\documentclass[a4paper,11pt]{article}
\pdfoutput=1 % if your are submitting a pdflatex (i.e. if you have
\usepackage{jheppub} % for details on the use of the package, please
\usepackage{subfigure}
\usepackage{xcolor}
\usepackage[T1]{fontenc} % if needed
\usepackage{xspace}
\usepackage{multirow}
\usepackage{hyperref}
\usepackage{slashed}
\usepackage{amsmath}
\usepackage{mathtools,amssymb,mathrsfs,tikz,dsfont}

\usepackage[utf8]{inputenc}

\newcommand{\Z}{{\mathds Z}}

\def\GeV{\ifmmode {\mathrm{\ Ge\kern -0.1em V}}\else \textrm{Ge\kern -0.1em V}\fi}%
\def\GeV{\ifmmode {\mathrm{\ Ge\kern -0.1em V}}\else \textrm{Ge\kern -0.1em V}\fi}%

\title{\boldmath Monopoles and Fermions in the Standard Model}

\author{Valentin V. Khoze}

\affiliation{IPPP, Department of Physics, Durham University, Durham DH1 3LE, UK}

\emailAdd{valya.khoze@durham.ac.uk}
\abstract{
We consider all magnetic monopoles that could have settled in the Standard Model after descending from a generic microscopic theory. These monopoles  have Standard Model quantum numbers, are stable, and we also require that their magnetic fluxes are consistent with the electroweak symmetry breaking. Scattering processes involving quarks, leptons and protons on these monopoles are studied using  partial waves decomposition. These processes in the lowest partial wave are known to be unsuppressed by the monopole mass  and are relevant for  monopole catalysis of proton decay. We provide estimates for scattering cross-sections  and investigate and confirm the applicability of the twisted sector approach to scattering processes on these Standard Model monopoles. We find that the SM monopole catalysis processes are universal and model-independent.}

\begin{document}
%\preprint{IPPP/24/XX}

\maketitle

\flushbottom
%\newpage

%%%%%%%%%%%%%%%%%%%%%%%%%%%%%%%%%%%%%%%%%%%%%%%

\section{\label{Sec:Intro}Introduction}

Since their theoretical inception~\cite{Dirac:1931kp}, 
magnetic monopoles have fascinated particle physics community thanks to their
unique reach and links with distinct theoretical physics areas and applications. These include an explanation for quantisation of the electric charge,  applications to cosmology, baryon and lepton number violation, potential applications in cosmic ray physics, also strong dynamics, confinement and duality. 

What is most relevant for this paper is the subtle role monopoles play in scattering processes with electrically charged 
particles. 
In particular, if a proton finds a monopole to scatter on in a head-on collision with kinetic energy of the order of few GeV,
the proton would decay. The characteristic scale for this reaction is set by strong interactions, the proton decay is rapid. 
At the parton level there are  Callan--Rubakov scattering processes of the type~\cite{Rubakov:1981rg,Callan:1982ah},
\begin{equation}
\label{eq:m-cat-p}
 u^1 \,+\,u^2 \,+\,  M \,\to\, \bar{d}^{3} \,+\, \bar{e} \,+\,  M
\end{equation}  
which result in the monopole catalysis of proton decay.
 Since the process is insensitive to the monopole mass, 
it is unsuppressed~\cite{Rubakov:1981rg,Rubakov:1982fp,Callan:1982ah,Callan:1982au}
by large scales of the microscopic UV theory.

If we consider a single fermion in the initial state scattering on a monopole in the lowest partial wave, the outcome is even more surprising and it 
appears to necessarily involve final states with fractional fermion numbers~\cite{Callan:1982au,Callan:1983tm},
\begin{equation}
 e\, +\, M \,\to\,  \frac{1}{2}\left(e\,\bar u_{1}\, \bar u_{2}\,\bar d_{3} \right)
 \,+\, M
\end{equation} 
The final state  consists of four semitons, each with a fermion number 1/2. 
These states cannot be described by the usual asymptotic Fock states in perturbation theory, they exist only in the massless limit and disintegrate 
at distances further away from the monopole where mass effects become important.

In a recent paper the authors of~\cite{vanBeest:2023dbu} provided an explanation of what the final states Fock space is and why it is different from the standard one.
The Fock space of (massless) final states lives in a twisted sector related to the perturbative one by an action of a discrete symmetry.
According to~\cite{vanBeest:2023dbu}, the outgoing radiation carries {\it integer} fermion numbers, while the fractional charges 
become the property of the vacuum state.
\medskip

Our main motivation is to apply and investigate this construction to the Standard Model. 
It was pointed out in Ref.~\cite{vanBeest:2023mbs} that the chiral nature of fermion representations creates an obstacle for 
applying the twisted sector formalism~\cite{vanBeest:2023dbu} for all but the simplest monopoles in the Standard Model.
We find that these restrictions are circumvented when we require that the monopoles respect the electroweak symmetry breaking (EWSB). 
One could of course argue that at very short distances, near the GUT-scale monopole core, the electroweak scale masses are negligible 
and one can ignore the effects of EWSB. 
But at such short scales the concept of asymptotic states is suspect and without good asymptotic states we cannot talk about scattering.
Our conclusion is that for all of the SM monopoles in the theory after EWSB, the application of the twisted sector method works out well and there
are no apparent obstacles, except for the usual interpretational issues of precisely how the twisted sector vacuum or the semitons disintegrate when fermion masses become important.

The second aim of the paper is to link these formal developments with analytic expressions 
for scattering amplitudes and to provide estimates for the cross-sections that can be used in phenomenological studies.   

\medskip 
The paper is organised as follows. In section~\ref{Sec:QED} we consider fermions scattering on a monopole in pure QED and 
summarise key facts about partial-waves and the Kazama, Yang and Goldhaber (KYG) result~\cite{Kazama:1976fm} for the cross-section.

In section~\ref{sec:3} we classify monopoles in the Standard Model before and after EWSB.
The monopoles are fully characterised by the magnetic fluxes of the SM gauge groups
which do not depend on the specific UV completion of the SM or details of the SM embedding.
Tables~\ref{Tab:Mpclass} and \ref{Tab:Mpcpr} provide this classification of SM monopoles and also map it to the monopoles in the minimal GUT theory.

In section~\ref{Sec:j_0-rules} we discuss the selection rules for the $j_0$-wave scattering where helicity-flip and $(B+L)$-violating processes occur. Section~\ref{Sec:minM} contains a detailed discussion of scattering on the minimal GUT monopole $M_1$.
In section~\ref{Sec:inter1} we turn what we have leaned so far into phenomenological cross-sections and provide estimates for different types of processes.
This section contains some of our main results, and these will be re-used also in the following sections considering the rest of the SM monopoles.

Section~\ref{Sec:ewM}
considers $M_3$ and  $M_3^\prime$ monopoles which carry no chromomagnetic fluxes.
We find in sections~\ref{Sec:ewM}-\ref{Sec:otherM} that for all SM monopoles, 
$M_1, M_2^{\,\prime},\ldots\,M_6^{\,\prime}$,
that respect EWSB
there are no obstacles in describing fermion scattering on any of the SM monopoles after EWSB which is one of our main observations.
Section~\ref{Sec:concl} presents a summary and conclusions.

%%%
\bigskip

\centerline{\it A note on conventions}
\medskip
%%%

In our notation the anti-particle of a left-handed fermion $f_{L\,\alpha}$ is described by the fermion field of the right-handed chirality, 
$(\overline{f_L})_R^{\,\, \dot{\alpha}}$ and vice versa, for example,
\begin{eqnarray}
 (u_{L}^{\,i})^\dagger \,=\, ( \overline{u_L})_{R\,i}
\qquad{\rm and}\qquad
 (e_R)^\dagger \,=\, ( \overline{e_R})_{L}
 \end{eqnarray}
The raised index $i$  indicates that the field is in the fundamental representation (in this case of $SU(3)_c$);  after hermitian conjugation symmetry group representations change to the conjugate representations which is represented by the lowered index.

\medskip

 %%%%%%%%%%%%%%%%%%%%%%%%
\section{Dirac monopoles and fermion-monopole scattering in QED}
\label{Sec:QED}
%%%%%%%%%%%%%%%%%%%%%%%%

In this section we summarise some of the well-known facts related to magnetic monopoles and their interactions with electrically charged fermions in QED. This is a warm-up to the scattering of fermions on monopoles in the Standard Model (SM).

Magnetic monopoles in QED are described by the Dirac gauge field configuration~\cite{Dirac:1931kp},
 \begin{equation}
{\cal A}_\mu^{\,\, (q_m){\rm Dirac}}\,=\, \frac{q_m}{2}\, (1-\cos \theta) \, \partial_\mu \phi\,,
\label{eq:ADirqm}
\end{equation}
where $q_m$ is the monopole's $U(1)$ magnetic charge and we have used conventional spherical polar angles $\theta$ and $\phi$ for the monopole placed at the origin $r=0$. To facilitate subsequent applications to the SM, the gauge fields are rescaled by the corresponding coupling constant(s), ${\cal A}_\mu \,=\, g A_\mu$, 
in our case $g=e$.
Given a matter field $\psi_{q_e}$ with an electric charge $q_e$, the Dirac quantisation condition requires that the quantity, 
\begin{equation}
q_J \,:=\, \frac{q_e q_m}{2}  \,\in \, \frac{1}{2} \Z \,,
\label{eq:qJqed}
\end{equation}
is quantised in half-integer units, and each pair of electric and magnetic states in QED with charges $(q_e,q_m)$ is characterised by a half-integer parameter
$q_J$.

Scattering amplitudes of electrically charged particles in QED on magnetic monopoles, are conventionally described the partial-wave decomposition 
 approach pioneered of~\cite{Kazama:1976fm}.
In this approach the wave-functions of charged states are expanded in the basis of generalised spherical harmonics in the monopole background~\cite{Wu:1976ge}.
These spherical harmonics are the eigenfunctions of ${\mathbf J}^2$ and $J_z$, where  ${\mathbf J}$ is the conserved total angular momentum operator,
 \begin{equation}
 {\mathbf J}\,=\, {\bf r}\times ({\bf p}- q_e{\mathbf {\cal A}})\,+\, {\mathbf S} \,-\, q_J \, \hat{\bf r}\,.
 \label{eq:Jdef}
\end{equation}
The first term on the right hand side is the orbital momentum momentum in the background ${\cal A}_\mu$ of the Dirac monopole~\eqref{eq:ADirqm}, the second term is the spin of the electrically charged field, for Weyl fermions it is given by Pauli matrices, ${\mathbf S}={\mathbf \sigma}/2$. Notably, the final term in~\eqref{eq:Jdef} shifts the overall expression by the unit radius-vector times $q_J$. 
This means that even at infinite separations between the electric charges and the monopole, there is a non-vanishing angular momentum contribution$~\propto q_J$, and these states are pairwise entangled~\cite{Zwanziger:1972sx}. The half-integer-valued parameter $q_J$ that characterises this entanglement
is also the effective charge quantifying the strength of the electric-magnetic interaction in the quantum-mechanical scattering amplitudes and cross-sections that was constructed in~\cite{Kazama:1976fm} and is shown below
in Eq.~\eqref{eq:sigKYG} .

Wave-functions of fermion states in QED, that are being scattered on a monopole, are expanded in the basis of 
the generalised spherical functions, $\psi(x) = \sum_{j,s} \psi_{j,s}(t,r)\, f_{j,s}(\Omega)$, which are simultaneous eigenfunctions of ${\mathbf J}^2$ and $J_z$ operators,
\begin{equation}
{\mathbf J}^2 \,f_{j,s}(\Omega)\,=\, j(j+1)  \,f_{j,s}(\Omega)\,\,, \quad 
J_z \,f_{j,s}(\Omega)\,=\, s \,f_{j,s}(\Omega)\,,
\end{equation}
and, as such, are labelled by two indices, $j=j_0, j_0+1, \ldots, \infty $ and $s= -j, -j+1,\ldots j-1,j$, and so are the the wave-functions $\psi_{j,s}(t,r).$
For a given $j$, a 1-particle electric state $\psi_{j={\rm fixed},s}(t,r)$ forms a $(2j+1)$-dimensional representation of the $SU(2)_J$ rotation group of the total angular momentum~\eqref{eq:Jdef}.
The minimal value of $j$ (i.e. the lowest partial-wave) is $j_0$, which for a fermion of electric charge $q_e$ scattered on a monopole of magnetic charge $q_m$ is,
\begin{equation}
j_0 = |q_J| -1/2\,.
\label{eq:j0def}
\end{equation}
Kazama, Yang and Goldhaber (KYG)~\cite{Kazama:1976fm} computed scattering amplitudes of fermions on a static (i.e. infinitely heavy)
Dirac monopole in non-relativistic quantum mechanics (QM). In the lowest partial wave, $j_0$, they found that the only processes allowed involved a $L\leftrightarrow R$ helicity flip of the scattered fermion, specifically:
\begin{align}
	j=j_0: \quad \left\{\begin{aligned}
\,\,q_J >0:\qquad 		& \psi_{L\,j_0,s}\,+\,M \,\to\, \psi_{R\,j_0,s}\,+\,M  \\
\,\,q_J <0:\qquad 		& \psi_{R\,j_0,s}\,+\,M \,\to\, \psi_{L\,j_0,s}\,+\,M  \,,
\label{eq:hfKYG}
\end{aligned}
 \right.
\end{align}
while for all higher partial-waves the fermion helicity  does not change from the incoming to the outgoing fermion,
\begin{equation}
j\ge j_0+1: \quad 		\psi_{L\,j,s}\,+\,M \,\to\, \psi_{L\,j,s}\,+\,M\ \quad {\rm and} \quad
\psi_{R\,j,s}\,+\,M \,\to\, \psi_{R\,j,s}\,+\,M.
\label{eq:hnfKYG}
\end{equation}

A simple way to understand the selection rule~\eqref{eq:hfKYG} is to assemble a pair of Weyl fermions of the same QED electric charge into a single Dirac fermion and notice that $j_0$-wave solution of the Dirac equation in the monopole background has the form,
\begin{equation}
\psi_{D}\,=\, \begin{pmatrix} \psi_{L}\\ \psi_{R} \end{pmatrix}
 \,=\,
\frac{1}{r} \begin{pmatrix} \psi_s(t+r)\\ \tilde\psi_s(t-r) \end{pmatrix} f_{j_0,s} (\Omega) \,,\qquad {\rm for}\, q_J >0\,.
\label{eq:DirLR}
\end{equation}
This means that the left-handed Weyl fermion with positive $q_J$ in the lowest $j_0$-wave can be only incoming and that
the corresponding right-handed fermion has only the outgoing component when scattered on a monopole.\footnote{If $q_J <0$, the upper and lower entries on the right-hand side of~\eqref{eq:DirLR} interchange, and the right-handed fermions become incoming and vice versa.} Hence the helicities of the incoming and outgoing fermions have to flip in accordance with the KYG selection rule~\eqref{eq:hfKYG}. 

\medskip

The $j_0$~partial-wave differential cross-section for both processes in~\eqref{eq:hfKYG} is~\cite{Kazama:1976fm},\footnote{Note that we assume that
 incoming fermions have fixed polarisation.
The result quoted in~\cite{Kazama:1976fm} has an extra factor of $1/2$ because they take their cross-section to be unpolarised. Our expressions 
in~\eqref{eq:sigKYG} and~\eqref{eq:mgRuth} are for polarised incoming fermions.}
\begin{equation}
\frac{d\sigma_{j_0}}{d\Omega}\,\,=\, \frac{q_J^2}{p^2}\, \left(\sin \theta/2 \right)^{4|q_J|-2}
\label{eq:sigKYG}
\end{equation}
where $p$ is the 3-momentum and $\theta$ is the polar angle of the scattered fermion in the laboratory frame where the monopole is at rest.

Cross-sections for all higher-partial-wave processes~\eqref{eq:hfKYG} have also been computed, summed over $j$ and tabulated in~\cite{Kazama:1976fm}.
In distinction with~\eqref{eq:hfKYG}, the higher-$j$ cross-sections have a Rutherford-type behaviour $\sim 1/\left(\sin \theta/2 \right)^{4}$ 
at small angles, 
\begin{equation}
\frac{d\sigma_{j>j_0}}{d\Omega}\,\,\sim\,\,\frac{q_J^2}{2 p^2}\frac{1}{\left(\sin \theta/2 \right)^{4}}
\label{eq:mgRuth}
\end{equation}
see Ref.~\cite{Kazama:1976fm} for details of the angular dependence for general values of $\theta$. We also note that the actual (unpolarised) Rutherford cross-section for all electrically charged particles, e.g. electrons scattered on muons,
\begin{equation}
\frac{d\sigma_{R}}{d\Omega}\,=\,\,\frac{\alpha^2}{4 p^2\beta^2}\, \frac{1}{\left(\sin \theta/2 \right)^{4}}
\label{eq:Ruth}
\end{equation}
 contains an additional $1/\beta^2$ enhancement factor relative to~\eqref{eq:mgRuth}. The reason it is absent on the right hand side of~\eqref{eq:mgRuth} is that the Lorentz force between the electron and the monopole is also proportional to $\beta=v/c$ and the resulting factors of $\beta^2$ cancel in the numerator and the denominator.

The coupling constant factor comparison between the electric-electric Rutherford scattering~\eqref{eq:Ruth} and the electric-magnetic KYG formulae~\eqref{eq:mgRuth} relates the fine structure constant squared $\alpha^2$ to
the pairwise helicity factor $q_J^2$, which is what it should be since,
\begin{equation}
q_J^2 \,=\, \alpha \, \alpha_M,
\end{equation}
where the magnetic coupling constant is
\begin{equation}
\alpha_M\,=\, \frac{g_M^2}{4\pi}\,, \quad {\rm and} \quad g_M \,:=\, q_J \, \frac{4\pi}{e}
\end{equation}

\medskip

Note that the KYG result~\eqref{eq:sigKYG} must also be correct in a fully relativistic theory for elastic $2\to2$ processes in the limit where the monopole mass $M_M \gg m_{\psi}$. This is indeed the case, as was demonstrated in~\cite{Csaki:2020inw}, using the pairwise helicity formalism for electric-magnetic scattering amplitudes developed by these authors, and the requirement that the helicity-flip amplitudes for~\eqref{eq:hfKYG} should saturate unitarity of the $S$-matrix in the $2\to2$ sector.

%%%
\section{Monopoles in the Standard Model}
\label{sec:3}
%%%

There are two points that we would like to note: 
\medskip

{\rm 1.}  All the monopoles that may or may not be present in the Standard Model must descend from some underlying theory in the UV.   
Since only finite-energy non-singular field configurations are admissible as monopole avatars, their existence necessarily requires a UV extension of the SM. 

{\rm 2.} Fortunately, neither the knowledge of the specific UV theory selection nor the technical details of the SM embedding are needed in order 
to classify all possible magnetic monopole states in the Standard Model. 
The monopoles are fully characterised by the quantum numbers of magnetic fluxes of the SM gauge groups, their topological magnetic charge, and the requirement that these configurations are stable within the Standard Model. This information is contained in the Standard Model itself.

\medskip
\medskip
 
The Dirac monopole configuration in QED~\eqref{eq:ADirqm} is singular, its energy density diverges at $r\to 0$ and,
if taken literally, would result in an infinite-mass state. It should be seen instead as a low-energy effective theory representation of a finite-mass monopole solution of the underlying UV theory. For an appropriate UV completion of QED that would support such solutions, the monopole would have a core of radius $R_M\sim 1/{\cal M}$ 
given by the inverse mass scale of the UV theory with the energy density convergent inside the core. Principal examples of finite-energy monopoles are the
't~Hooft-Polyakov monopole solutions~\cite{tHooft:1974kcl,Polyakov:1974ek} in a non-Abelian gauge theory with a Higgs field in the adjoint representation, and the monopoles~\cite{Kim:1999bd} of the Dirac-Born-Infeld (DBI) extensions of QED 
and/or other sectors of the SM.
't Hooft-Polyakov monopoles exist in all UV extensions of the SM where the $U(1)_Y$ gauge factor is compact, i.e. originates from a non-Abelian gauge factor. This occurs in all Grand Unified Theories (GUT) settings. The 't~Hooft-Polyakov monopole mass, $M_M \sim {\cal M}/\alpha$, and the non-Abelian core,  
$R_M\sim 1/{\cal M}$, are set by the masses ${\cal M}$ of the heavy GUT vector bosons. Other than GUT non-Abelian embeddings of $U(1)_Y$, e.g. into a separate $SU(2)_R$ group,  also give rise to 't Hooft-Polyakov monopoles. The second category where finite-mass, regular-core monopole solutions are present~\cite{Kim:1999bd} is when the $U(1)_Y$
theory is embedded into a Dirac-Born-Infeld (DBI) formulation.
%\begin{equation}
%{\cal L}_{\rm DBI}\,=\,
%\beta^2_{\rm DBI} \left(1-\sqrt{-{\rm det} (\eta_{\mu\nu} +(1/\beta_{\rm DBI}) F_{\mu\nu})   }
%\right)
%\end{equation}
In this case genuine regular monopoles also known to exist and their core is set by the DBI mass scale.
In all these scenarios, 
%the regular monopoles of the microscopic theory approach Dirac monopole configurations in the decoupling limit when ${\cal M}\to \infty$, or equivalently, at large distances $r \gg R_M$.
%
%\medskip
%At 
at distances greater than the monopole core $r\gg R_M$  any spherically symmetric monopole configuration is gauge-equivalent to a Dirac monopole times a constant matrix from the Lie algebra of the SM gauge 
group~\cite{Coleman:1982cx,Preskill:1984gd},\footnote{In our notation the gauge field is Lie algebra-valued, ${\cal A}_\mu = g_s A_\mu^a T^a_{c} +  g_w A_\mu^b T^b_{L}  + g' B_\mu$, where $g_s$, $g_w$ and $g'$ are the SM gauge coupling constants, and $T^a_{c}$ and $T^b_{L}$ are the generators of $SU(3)_c$ and $SU(2)_L$.}
\begin{equation}
{\cal A}_\mu \,=\,\, Q_{\rm mg} \, {\cal A}_\mu^{\,{\rm Dirac}}\,=\,\, Q_{\rm mg} \,\, \frac{1}{2}\, (1-\cos \theta) \, \partial_\mu \phi
\,, 
\label{eq:Amumon}
\end{equation}
where ${\cal A}_\mu^{\,{\rm Dirac}}$ is the Dirac monopole and 
$Q_{\rm mg}$ is a constant diagonal matrix from the Cartan subalgebra of the gauge group that is constrained by the Dirac quantisation condition. 

It is well-known that monopoles are sensitive to the global structure of the SM gauge group~\cite{Tong:2017oea}.\footnote{For a recent discussion of possible
applications of the $G_{SM}$ global structure to SM phenomenology see e.g.~\cite{Alonso:2024pmq,Li:2024nuo}.}
In what follows, we will concentrate on the SM group with the largest allowed center ($\Z_p=\Z_6$) that is consistent with the SM filed content 
and which allows for the largest selection of monopole states~\cite{Tong:2017oea},
\begin{equation}
G_{SM} \,=\, \frac{SU(3)_c \times SU(2)_L \times U(1)_Y}{\Z_6}\,.
\label{eq:G6}
\end{equation}

\noindent In this case the magnetic flux matrix $Q_{\rm mg}$ is given by~\cite{vanBeest:2023mbs},
\begin{equation}
G_{SM}:\quad  Q_{\rm mg}\,=\, \begin{pmatrix} n_1-\frac{n_1+n_2}{3}& &\\ & n_2 -\frac{n_1+n_2}{3}&\\ & & -\frac{n_1+n_2}{3}
\end{pmatrix} 
\oplus
\begin{pmatrix} \frac{m}{2} & \\ &-\frac{m}{2}
\end{pmatrix} 
\oplus \, p\,,
\label{eq:FmgSM}
\end{equation}
and is characterised by four integers $(p\,; n_1, n_2, m) \in \Z$ that account for the  Abelian and the non-Abelian magnetic fluxes of the monopole.
Importantly, only the Abelian flux $p$ is protected by topology: it is the winding number of the closed contour $S^1$ around the Dirac string
 to the $U(1)_Y$ group manifold, $p= \pi_1(U(1)_Y) \in \Z$ \cite{Lubkin:1963zz}. The non-Abelian magnetic flux numbers,
 $(n_1, n_2, m)$, on the other hand, can change by monopole emitting massless non-Abelian SM gauge fields~\cite{Brandt:1979kk,Coleman:1982cx}.
The authors of~\cite{vanBeest:2023mbs} pointed out that  for each value of $p$ there is precisely one monopole, $M_p$, that is stable against decay into massless $SU(2)_L$ vector bosons and $SU(3)_c$ gluons of \eqref{eq:G6}.
We thus have 6 elementary monopoles $M_1, \ldots, M_6$ that carry SM magnetic fluxes (i.e. interact with SM particles)  and are perturbatively stable in the massless SM  i.e. {\it before} EWSB takes place.
 The allowed values of  $(p\,; n_1, n_2, m)$  and the corresponding $M_p$ monopoles are~\cite{vanBeest:2023mbs}:
\begin{align}
\begin{aligned}
\,\,M_1:\, (1;1,1,1)\qquad 	 & \,\,M_2:\, (2; 1,0,0)\qquad     \,\,M_3:\, (3; 0,0,1)\qquad    \\
\,\,M_4:\, (4;1,1,0)\qquad 	 & \,\,M_5:\, (5; 1,0,1)\qquad     \,\,M_6:\, (6; 0,0,0)\qquad
\end{aligned}
 \label{eq:FmgSMstab}
\end{align}
These monopoles along with their flux matrices $Q_{\rm mg}$  are presented in Table~\ref{Tab:Mpclass}.
%
%In displaying magnetic flux matrices in~Table~\ref{Tab:Mpclass} we 
Note that there is a gauge freedom to re-arrange the order of the diagonal elements in the matrix $Q_{\rm mg}$ in~\eqref{eq:FmgSM} which was was fixed in~\cite{vanBeest:2023mbs} by choosing $n_1\ge n_2 \ge 0$ and $m\ge 0$. This selection was also assumed in~\eqref{eq:FmgSMstab}. 
In~Table~\ref{Tab:Mpclass} we use a more symmetric selection for the $SU(3)_c$ flux matrix $Q_{{\rm mg}\, c}^{(n_1,n_2)}$, where $Q_{{\rm mg}\, c}^{(1,0)}$ for $(n_1,n_2)=(1,0)$ is replaced by  $-Q_{{\rm mg}\, c}^{(1,1)}$
which is gauge  equivalent, 
\begin{eqnarray}
Q_{{\rm mg}\, c}^{(1,0)} \,=\, 
\begin{pmatrix} \frac{2}{3}& &\\ &- \frac{1}{3}&\\ & & -\frac{1}{3}\,
\end{pmatrix}  \,\,\equiv\,\,%\,\,\leftrightarrow\,\,
\begin{pmatrix} -\frac{1}{3}& &\\ & -\frac{1}{3}&\\ & & \frac{2}{3}\,
\end{pmatrix} \,=\, - \, Q_{{\rm mg}\, c}^{(1,1)}
\end{eqnarray}

 \begin{table}[ht]
\begin{equation} \nonumber
\begin{array}{lc|c|c}
 \text{ Monopoles before EWSB} && \text{general}\,\,\, Q_{{\rm mg}\,\,SM}  &\,Q_{{\rm mg}\,\,SU(5)_{GUT}} \,
\\  \hline  & & & \\
\quad {M_1}  &&\quad \begin{pmatrix} \frac{1}{3}& &\\ & \frac{1}{3}&\\ & & -\frac{2}{3}\,
\end{pmatrix} 
\oplus
\begin{pmatrix} \frac{1}{2} & \\ &-\frac{1}{2}
\end{pmatrix} 
\oplus \, 1\quad
& \quad T_c^8\,+\,T_L^3\,+\, Y \,
\\ & & & 
\\ \hline  & & & \\
\quad {M_2}  &&\quad \begin{pmatrix} -\frac{1}{3}& &\\ &-\frac{1}{3}&\\ & & \frac{2}{3}\,
\end{pmatrix} 
\oplus
\begin{pmatrix} 0 &  \\  & 0
\end{pmatrix} 
\oplus \, 2 \quad
& \quad -T_c^8\,+\,2 Y \,
\\ & & & \\ \hline  & & & \\
%%%
\quad {M_3}  &&\quad \begin{pmatrix} 0& &\\ &0&\\ & & 0 \,
\end{pmatrix} 
\oplus
\begin{pmatrix} \frac{1}{2} & \\ &-\frac{1}{2}
\end{pmatrix} 
\oplus \, 3 \quad
& \quad T_L^3\,+\,3 Y \,
\\ & & & \\ \hline  & & & \\
%%%%
\quad {M_4}  &&\quad \begin{pmatrix} \frac{1}{3}& &\\ &\frac{1}{3}&\\ & & -\frac{2}{3}\,
\end{pmatrix} 
\oplus
\begin{pmatrix}  0 & \\ & 0
\end{pmatrix} 
\oplus \, 4 \quad
& \quad T_c^8\,+\,4 Y \,
\\ & & & \\ \hline  & & & \\
%%%%%
\quad {M_5}  &&\quad \begin{pmatrix} -\frac{1}{3}& &\\ &-\frac{1}{3}&\\ & & \frac{2}{3}\,
\end{pmatrix} 
\oplus
\begin{pmatrix} \frac{1}{2} & \\ &-\frac{1}{2}
\end{pmatrix} 
\oplus \, 5 \quad
& \quad -T_c^8\,+\,T_L^3\,+\, 5Y \,
\\ & & & \\ \hline  & & & \\
%%%%%%
\quad {M_6}  &&\quad \begin{pmatrix} 0& &\\ &0&\\ & &0\,
\end{pmatrix} 
\oplus
\begin{pmatrix} 0 &  \\  & 0
\end{pmatrix} 
\oplus \, 6 \quad
& \quad 6 Y \,
\\ & & & \\ \hline  %& & & \\
\end{array}
\end{equation}
\caption{Monopoles in the SM before EWSB.  The second column uses general universal notation for SM fluxes~\eqref{eq:FmgSM}-\eqref{eq:FmgSMstab}. The third column proves their equivalence 
to the monopoles realised in the minimal $SU(5)_{GUT}$ settings~\eqref{eq:Qsu5}-\eqref{eq:Ysu5}.}
\label{Tab:Mpclass}
\end{table}

\medskip 

It is worthwhile to point out  that the general monopole classification described above %in~\eqref{eq:FmgSM}-\eqref{eq:FmgSMstab} 
is actually {\it equivalent} to what can be obtained from the {\it minimal} embedding of the SM into the $SU(5)_{GUT}$ theory. 
It does not matter whether the minimal $SU(5)_{GUT}$ was the physically realised UV theory of the SM at any stage in the early universe. In so far as our goal is to identify all possible UV monopoles that are stable against radiative decay into massless SM degrees of freedom, and which are characterised by their magnetic fluxes of the low-energy theory~\eqref{eq:G6}, we can equally well use $SU(5)_{GUT}$ as formal book-keeping device for SM gauge and matter fields.
To see this we recall that for all spherically-symmetric $SU(5)_{GUT}$ monopoles their magnetic are of the general form~\cite{Gardner:1983uu}, 
\begin{equation}
Q_{{\rm mg}}\,=\, n_8 T_c^8\,+\, n_3 T_L^3\,+\, p Y\,,
\label{eq:Qsu5}
\end{equation}
where we introduce the conventional colour-, weak- and $Y$-hypercharge generators of the $SU(5)_{GUT}$,
\begin{eqnarray}
T_c^8 &=& {\rm diag}(1/3, 1/3, -2/3 ,0 ,0)\,,   \nonumber \\
T_L^3 &=& {\rm diag}(0,0,0,1/2,-1/2)\,,    \label{eq:Ysu5}  \\
Y &=& {\rm diag}(-1/3,-1/3,-1/3,1/2,1/2)\,, \nonumber
\end{eqnarray}
and in this notation the electric charge generator $Q_{\rm el}$ satisfies the usual Gell-Mann--Nishijima formula,
\begin{equation}
Q_{\rm el} \,=\, T_L^3 \,+\, Y\,=\, {\rm diag}(-1/3,-1/3,-1/3,1,0)\,. 
\end{equation}
There is a on-to-one correspondence between \eqref{eq:Qsu5} and the magnetic fluxes in the second column in the Table~\ref{Tab:Mpclass}
with $n_3=m$, $p=p$ and $n$ being equal to the coefficient of the $SU(3)_c$ matrix $Q_{{\rm mg}\, c}^{(1,1)}$.
We conclude that the monopoles resulting from~\eqref{eq:Qsu5} are identical to those specified by~\eqref{eq:FmgSM}-\eqref{eq:FmgSMstab}, as can be seen by comparing the second and the third column in~Table~\ref{Tab:Mpclass}.

 \begin{table}[t]
\begin{equation} \nonumber
\begin{array}{lc|c|c}
 \text{ Monopoles after EWSB} &&\text{general}\,\,\, Q_{{\rm mg}\,\,SM^\prime} &\,Q_{{\rm mg}\,\,SU(5)_{GUT}} \,
\\  \hline  & & & \\
\quad {M_1^{\,\prime}=M_1}  &&\quad 
\begin{pmatrix} \frac{1}{3}& &\\ & \frac{1}{3}&\\ & & -\frac{2}{3}\,
\end{pmatrix} 
\oplus
\begin{pmatrix} \frac{1}{2} & \\ &-\frac{1}{2}
\end{pmatrix} \,=\, 
\oplus \, 1\quad
& \quad T_c^8\,+\,Q_{\rm el} \,
\\ & & & \\ \hline  & & & \\
\quad M_2^{\,\prime}  &&\quad \begin{pmatrix} -\frac{1}{3}& &\\ &-\frac{1}{3}&\\ & & \frac{2}{3}\,
\end{pmatrix} 
\oplus
\begin{pmatrix} 1 &  \\  & -1
\end{pmatrix} 
\oplus \, 2 \quad
& \quad -T_c^8\,+\,2 Q_{\rm el}  \,
\\ & & & \\ \hline  & & & \\
%%%
\quad M_3^{\,\prime}  &&\quad \begin{pmatrix} 0& &\\ &0&\\ & & 0 \,
\end{pmatrix} 
\oplus
\begin{pmatrix} \frac{3}{2} & \\ &-\frac{3}{2}
\end{pmatrix} 
\oplus \, 3 \quad
& \quad 3 Q_{\rm el}  \,
\\ & & & \\ \hline  & & & \\
%%%%
\quad M_4^{\,\prime}  &&\quad \begin{pmatrix} \frac{1}{3}& &\\ &\frac{1}{3}&\\ & & -\frac{2}{3}\,
\end{pmatrix} 
\oplus
\begin{pmatrix}  2 & \\ &-2
\end{pmatrix} 
\oplus \, 4 \quad
& \quad T_c^8\,+\,4 Q_{\rm el} \,
\\ & & & \\ \hline  & & & \\
%%%%%
\quad M_5^{\,\prime}  &&\quad \begin{pmatrix} -\frac{1}{3}& &\\ &-\frac{1}{3}&\\ & & \frac{2}{3}\,
\end{pmatrix} 
\oplus
\begin{pmatrix} \frac{5}{2} & \\ &-\frac{5}{2}
\end{pmatrix} 
\oplus \, 5 \quad
& \quad -T_c^8\,+\, 5Q_{\rm el}  \,
\\ & & & \\ \hline  & & & \\
%%%%%%
\quad M_6^{\,\prime}  &&\quad \begin{pmatrix} 0& &\\ &0&\\ & &0\,
\end{pmatrix} 
\oplus
\begin{pmatrix} 3 &  \\  & -3
\end{pmatrix} 
\oplus \, 6 \quad
& \quad 6 Q_{\rm el}  \,
\\ & & & \\ \hline  %& & & \\
\end{array}
\end{equation}
\caption{Monopoles in the SM after EWSB~\eqref{eq:GEWSB}.  The second and third columns display SM fluxes in the general case~\eqref{eq:FmgSMpr} and in the equivalent to it $SU(5)_{GUT}$ settings.}
\label{Tab:Mpcpr}
\end{table}

\medskip

Note that there is a caveat to the general model-independent classification of SM monopoles presented above. Apart from the SM quantum numbers/fluxes, one can also ask more detailed questions about the monopole mass spectrum and whether all of the $M_p$ monopoles are stable against separation into pairs of  more elementary monopoles with lower values of $p$.
These are dynamical questions that do depend on the specifics of  
the UV-theory, especially when it contains a hierarchy of different UV-scales and Higgs fields contributing to the symmetry breaking patterns and monopole production. 
For example, it was demonstrated 
in~\cite{Gardner:1983uu} that in the minimal $SU(5)_{GUT}$ model all the monopoles in~Table~\ref{Tab:Mpclass}, except $M_5$, are stable against decay into  pairs of more elementary monopoles, while $M_5$ was unstable against a decay into a double and a triple monopole. These consideration a based on a computation of the monopole binding energy and are sensitive to the Higgs potential as well as the gauge degrees of freedom. The conclusions depend on a specific model. In the minimal GUT model the lightest monopole is the monopole with the minimal (single unit) of $U(1)_Y$ magnetic charge $M_1$, but 
as explained in section~5 of Ref.~\cite{Preskill:1984gd}, in $SO(10)_{\rm GUT}$ the doubly charged monopole $M_2$ can be much lighter than $M_1$. 
In general,
a UV theory with a complicated symmetry-breaking hierarchy would contain stable monopoles with widely different masses with the minimal $M_1$ monopole often being the heaviest. 
Keeping these points in mind, it is still the case that any magnetic monopole that is stable and has retained at least some of the SM magnetic fluxes  
(and thus can interact directly with SM particles) must given by one of the configurations in~Table~\ref{Tab:Mpclass}.\footnote{This statement also 
assumes that the monopole in the UV theory was spherically symmetric up to gauge transformations. This is required to justify~\eqref{eq:Amumon}.}

\medskip

We now have to account for the electroweak symmetry breaking (EWSB) of $G_{SM}$ to  
\begin{equation}
G_{SM}^{\,\prime} \,=\, \frac{SU(3)_c \times U(1)_{QED}}{\Z_3}\,,
\label{eq:GEWSB}
\end{equation}
induced by the SM Higgs vev. This is not merely an academic point as the EWSB modifies the long-range fields of SM monopoles 
by screening their non-Abelian magnetic fluxes associated with the now massive $Z^0$ bosons (i.e. the screening of magnetic neutral currents).
At distances $r > 1/M_{Z} \sim 1/ (100~{\rm GeV})$ monopole gauge field configurations cannot contain a non-vanishing component along the $Z^0$ vector boson direction, i.e. along the $T^3_L- Y$ generator. 

The requirement that the monopole gauge field has a vanishing $Z_\mu^0$ implies that $m$ must be set equal to $p$ in~\eqref{eq:FmgSM},
\begin{equation}
G_{SM}^{\,\prime}:\quad  Q_{\rm mg}\,=\, \begin{pmatrix} n_1-\frac{n_1+n_2}{3}& &\\ & n_2 -\frac{n_1+n_2}{3}&\\ & & -\frac{n_1+n_2}{3}
\end{pmatrix} 
\oplus
\begin{pmatrix} \frac{p}{2} & \\ &-\frac{p}{2}
\end{pmatrix} 
\oplus \, p\,,
\label{eq:FmgSMpr}
\end{equation}
This of course is not a matter of choice, but an automatic consequence of satisfying classical equations for the monopole in the presence of the SM Higgs field with the vev$\neq 0$. 

The monopole stability conditions~\eqref{eq:FmgSM} are modified accordingly. The non-Abelian $SU(2)_L$ bosons are no longer massless 
and the monopole with $m=p$ is now automatically stable in the  $SU(2)_L$  sector. At the same time, the decay mode into massless gluons are continued to be avoided by restricting $(n_1,n_2)$ in the same way as was done in~\eqref{eq:FmgSM}. 
In summary, after EWSB the monopoles surviving in the Standard Model are the modified monopoles $M_p^{\,\prime}$ which we detail in~Table~\ref{Tab:Mpcpr}. Note that only the minimal monopole configuration
remains the same, $M_1^{\,\prime} = M_1$, with all other monopoles having to adjust their $T^3_L$ flux to change the Abelian flux $pY$ into $p\,Q_{\rm el}$ in accordance with~\eqref{eq:FmgSMpr}, see the last column in~Table~\ref{Tab:Mpcpr}. The third column of the Table proves the equivalence of the full set of general 
stable $M_p^{\,\prime}$ monopoles with the monopoles in the minimal $SU(5)_{\rm GUT}$ realisation of monopoles after EWSB~\cite{Dokos:1979vu}.

%%%%%%%%%%%%%%%%%%%%%%%%
\section{Selection rules for $j_0$-wave scattering}
\label{Sec:j_0-rules}
%%%%%%%%%%%%%%%%%%%%%%%%

In the subsequent sections we will study interactions  of the SM monopoles $M_p$ and $M_p^{\,\,\prime}$ with SM fermions before and after the electroweak symmetry breaking. Most interesting are the scattering amplitudes in the lowest partial wave $j_0$ which can generate baryon and lepton number $(B+L)$-violation and lead to unsuppressed proton decay and proton generation processes catalysed by magnetic monopoles~\cite{Rubakov:1981rg,Rubakov:1982fp,Callan:1982ah,Callan:1982au}.
As in the case of QED, the long-range interactions of SM monopole with charged matter fields are characterised by the total angular momentum operator 
${\mathbf J}$ given in~\eqref{eq:Jdef}. The pairwise helicity variable $ q_J$ can be determined for each (monopole,~fermion)-pair by generalising the QED 
expression~\eqref{eq:qJqed} to the full non-Abelian SM case by summing over the strong, weak and $Y$-hypercharge pairs of the corresponding electric and magnetic charges.
Specifically, for a SM fermion $\psi_{(\boldsymbol3,\boldsymbol2)_Y}$  (such as the quark doublet $q_L$  in Table~\ref{Tab:SMfields})
transforming in the fundamental representations of $SU(3)_c$ and $SU(3)_L$ with the hypercharge $Y$, and the monopole with the magnetic flux matrix in~\eqref{eq:FmgSM} that we abbreviate as,
\begin{equation}
Q_{\rm mg} \,=\, Q_{{\rm mg}\,c} \oplus Q_{{\rm mg}\,L} \oplus p\,,
\end{equation}
the pairwise helicity is given by 
\begin{equation}
q_J \, =\, \frac{1}{2} \left(q_{{\rm mg}\,c} + q_{{\rm mg}\,L}\, +Yp  \right) \,,
\label{eq:qJSMgen}
\end{equation}
where the magnetic charges $q_{{\rm mg}\,c} $ and $q_{{\rm mg}\,L}$ are the eigenvalues of the non-Abelian magnetic matrices $Q_{{\rm mg}\,c}$ and $Q_{{\rm mg}\,L}$ %when acting on the fermion $\psi_{(\boldsymbol3,\boldsymbol2)_Y}$, 
and we have used the fact that non-Abelian electric charges are all =1 for the field in the fundamental representation, while the $U(1)_Y$ electric charge of $\psi_{(\boldsymbol3,\boldsymbol2)_Y}$ is $Y$.

\begin{table}[t]
    \centering
\begin{tabular}{c|c|c|c|c|c|c}
     & $q_L$& $u_R$& $d_R$ & $\ell_L$ & $e_R$ & $H$\\ \hline
    $SU(3)_c$ & $\boldsymbol3$ & $\boldsymbol3$ & $\boldsymbol3$ & $\boldsymbol1$ &$\boldsymbol1$& $\boldsymbol1$\\
    $SU(2)_L$ &$\boldsymbol2$ & $\boldsymbol1$ & $\boldsymbol1$& $\boldsymbol2$ & $\boldsymbol1$&$\boldsymbol2$\\
    $ Y$ & $\frac{1}{6}$ & $\frac{2}{3}$ &$-\frac{1}{3}$ &$-\frac{1}{2}$ &$-1$&$\frac{1}{2}$ 
\end{tabular}
\caption{SM matter fields, their charges and representations}
    \label{Tab:SMfields}
\end{table}

For a SM fermion that is a singlet under $SU(3)_c$ or $SU(3)_L$ or both, the corresponding non-Abelian magnetic charge $q_{{\rm mg}} $
on the right hand side of~\eqref{eq:qJSMgen} would be zero.

\medskip 

Since the magnetic flux matrices $Q_{\rm mg}$ and the electric hypercharges of the SM matter fields are consistent with Dirac quantisation conditions, the 
values  of $q_J$ on the right hand side of~\eqref{eq:qJSMgen}, evaluated using Table~\ref{Tab:SMfields},
will be automatically half-integer-valued. This can be checked directly for each combination of fermions and SM monopoles -- and will also follow from our results below.

The sign of $q_J$ for each (monopole,~fermion)-pair determines whether a given Weyl fermion is an incoming or an outgoing state 
in its $j_0$-wave scattering, and the absolute value of $q_J$ tells us how the fermion transforms under the action of the $SU(2)_J$ rotation of the pair. The dimension of the $SU(2)_J$ representation is $2j_0+1\,=\, 2|q_J|$ according to~\eqref{eq:j0def}.
If $q_J$ vanishes for a particular (fermion,~monopole) pair, the fermion does not scatter on the monopole (the corresponding $T$-matrix is trivial) and the fermion is removed from the scattering states analysis.

\begin{table}[ht]
\begin{equation} \nonumber
\begin{array}{lc|ccc}
 &&\quad q_J &\,SU(2)_{J} & \,SU(N_f)\, 
 \\ \hline 
\psi\,\, \text{incoming:}  \quad
&\psi_{L/R}\, &\,\,(+/-)&{\boldsymbol{2|q_J|}}&{\boldsymbol{N_f}}
\\
\tilde\psi\,\, \text{outgoing:} \quad 
&\psi_{R/L}\, &\,\,(+/-)&{\boldsymbol{2|q_J|}}&{\boldsymbol{N_f}}
\label{eq:2to2}
\end{array}
\end{equation}
\caption{Incoming and outgoing states assignments for scattering in a vector-like theory in the lowest partial wave. Representations of the rotation group $SU(2)_J$ and an assumed flavour $SU(N_f)$ symmetry are shown.}
\label{Tab:M_genf}
\end{table}

In summary, we proceed as follows: first select a monopole $M_p$ or $M_p^{\,\prime}$.  Next compute pairwise helicities $q_J$ for all SM fermions and project on the subspace of fermions with non-vanishing $q_J$. 
If we can assemble all Weyl fermions with the same value of $q_j$ and in the same flavour representation 
into Dirac doublets -- as we have done in QED in~\eqref{eq:DirLR} -- we can assign the incoming $\psi$ and the outgoing $\tilde\psi$ states accordingly, and 
 find a simple $2\to2$ scattering helicity-flip process in the $j_0$-wave,
 \begin{equation}
 \psi^{\,I} +M\,\to\, \tilde\psi^{\,I}+M\,,
 \label{eq:flip1}
\end{equation}
where the index $I$ is the flavour index.
This is analogous to the process~\eqref{eq:hfKYG} inn QED. This scenario corresponds to a vector-like theory after the non-vanishing $q_J$-projection, and the corresponding selection rules are summarised in~Table~\ref{Tab:M_genf}.

Such simple flavour-symmetry-preserving fermion pairing $ \psi^{\,I} \leftrightarrow \tilde\psi^{\,I}$ indicated by the process~\eqref{eq:flip1},
 however, is not guaranteed and is often  impossible for SM monopoles $M_p$
as was emphasised in~\cite{vanBeest:2023mbs}. This is because in a general SM monopole background the resulting 
fermion representations are not vector-like.  We now want to check how this works for $M_p$ or $M_p^{\,\prime}$ monopoles in the SM before/after EWSB
in a case by case basis.

%%%%%%%%%%%%%%%%%%%%%%%%
\section{Minimal GUT monopole $M_1$}
\label{Sec:minM}
%%%%%%%%%%%%%%%%%%%%%%%%

The SM monopole with the minimal value of the topological charge $p$ (a.k.a. the 1-form charge) is the $p=1$ monopole $M_1$, also known as the minimal GUT monopole. As already noted, this monopole, shown in the top row in Tables~\ref{Tab:Mpclass} and \ref{Tab:Mpcpr}, has no
magnetic flux along the $Z_\mu^0 \sim T_L^3-Y$ direction. This particular monopole is the same before and after EWSB. 

To establish how $M_1$ interacts with the SM fermions we compute its pairwise helicity $q_J$ with each of the SM quarks and leptons using the formula~\eqref{eq:qJSMgen} along with the magnetic fluxes in Table~\ref{Tab:Mpclass} and the 
charges of fermions listed in Table~\ref{Tab:SMfields}. We find that the subset of SM fermions with a non-vanishing $q_J$ is given for a single generation  by 
$e_{L/R}$, $u_{L/R}^{1,2}$ and $d_{L/R}^3$ (where superscripts denote the $SU(3)_c$ colour indices of SM quarks surviving the $q_J\neq 0$ projection)
 and list their charges in Table~\ref{Tab:M1ferm} below. The incoming/outgoing status of the fermions is determined by the 
 sign of $q_J$ following the general rule of Table~\ref{Tab:M_genf}, and the dimension of the rotation group  $SU(2)_{J}$ representation is $2|q_J|$.
 Since all SM fermions interacting the the $M_1$ monopole turn out to be singlets of $SU(2)_{J}$, this rotation symmetry will play no further role in the $M_1$ case.

 \begin{table}[ht]
\begin{equation} \nonumber
\begin{array}{lc|cccc}
\quad {M_1} \text{ monopole} &&\quad q_J &\,SU(2)_{J} & \,SU(2)_c\, & \,Q_{el}\,
\\\hline
&\,\, u_L^{\,1,2}\,& +\frac{1}{2}& \boldsymbol1& \boldsymbol2&+\frac{2}{3}
\\
\psi\,\, \text{incoming:}  
&d_R^{\,3}\, &-\frac{1}{2}&\boldsymbol1&{\boldsymbol1}& -\frac{1}{3}
\\
&\ e_R\,& -\frac{1}{2} &\boldsymbol1 &\boldsymbol1&-1
 \\ \hline 
&\,\, u_R^{\,1,2}\,& +\frac{1}{2}& \boldsymbol1& \boldsymbol2&+\frac{2}{3}
\\
\tilde\psi\,\, \text{outgoing:}  
&d_L^{\,3}\, &-\frac{1}{2}&\boldsymbol1&{\boldsymbol1}& -\frac{1}{3}
\\
&\ e_L\,& -\frac{1}{2} &\boldsymbol1 &\boldsymbol1&-1
\end{array}
\end{equation}
\caption{SM fermions and their charges in the $j_0$ partial-wave scattering on the minimal $M_1$ monopole. We show their representations under the rotation group  $SU(2)_{J}$ and the $SU(2)_{c}$ subgroup of the $SU(3)_{c}$ is spanned by the $i=1,2$ colours of $u$-quarks; $Q_{el}$ denotes the electric charge.}
\label{Tab:M1ferm}
\end{table}

\subsection{$(B+L)$-conserving fermion-monopole scattering}
\label{sec:5.1}

It is a curious fact that even though the Standard Model reduces to QED in the IR after EWSB, the quintessential in QED helicity-flip fermion-monopole
processes~\eqref{eq:flip1} of KYG were often discarded in the subsequent literature.
For the minimal monopole $M_1$ interacting with the fermions listed in Table~\ref{Tab:M1ferm}
the KYG-type helicity-flip $j_0$-wave scatterings are,
%\begin{eqnarray}
% u_L^{\,1,2}\, +\, M_1 \,&\to&\, u_R^{\,1,2}\,+\, M_1 \nonumber\\
% d_R^{\,3}\, +\, M_1 \,&\to&\, d_L^{\,3}\,+\, M_1 \nonumber \\
% e_R\, +\, M_1 &\to& e_L\,+\, M_1 
 %\label{eq:flipM1s}
%\end{eqnarray}
%
\begin{align}
	\psi^{\,I} +M\,\to\, \tilde\psi^{\,I}+M\,: \quad %\left\{
	\begin{aligned}
\,\,&u_L^{\,1,2}\, +\, M_1 \,\to\, u_R^{\,1,2}\,+\, M_1   \\
\,\,&d_R^{\,3}\, +\, M_1 \,\to\, d_L^{\,3}\,+\, M_1     \\
\,\, &e_R\, +\, M_1 \,\to e_L\,+\, M_1   
\end{aligned}
% \right.
 \label{eq:flipM1s}
\end{align}
These processes conserve the electric and colour charge as well as $B$ and $L$.

    At sufficiently high energies the processes~\eqref{eq:flipM1s} are expected to be suppressed (see e.g.~\cite{Bais:1982hm,Preskill:1984gd,Rubakov:1988aq})
by small fermion masses in the case of leptons 
or by (not as small) values of quark condensates for quarks.
In the strictly massless limit, the amplitudes for these processes indeed have to vanish.
For example, for $e_R\, +\, M_1 \,\to e_L\,+\, M_1 $ the electron $e_R$
in the initial state does not couple to the weak-sector gauge fields, while the final state electron $e_L$ does. For all processes in~\eqref{eq:flipM1s} the initial and final state fermions have different
gauge charges of the $SU(2)_L$ sector and, in the absence of the SM Higgs vev, one cannot write down non-vanishing gauge-invariant expressions 
characterising these processes.\footnote{For example the operator $(\overline{e_R})_{L}^\alpha\, e_{L\alpha}^{\,\,r}$ whose condensate on the monopole would correspond to the 3rd process in \eqref{eq:flipM1s} has an un-contracted $SU(2)_L$ index $r$ and is not a gauge-singlet -- unless one would do the obvious thing and contract it with the SM Higgs field $H_r$ (or the Higgs vev) and the electrons Yukawa coupling. We use the all-outgoing conventions and $\alpha=1,2$ denotes the spinor index for the left-handed chiral fermions.}

In reality the SM Yukawa interactions, and consequentially non-zero fermion masses, as well as dynamical effects of strong interactions
reduce the underlying symmetry of the problem to that of Table~\ref{Tab:M1ferm} and the scattering processes~\eqref{eq:flipM1s} will occur.  In fact, as we will see in section~\ref{Sec:inter1}, the cross-sections for these processes are large in the IR, 
and are being enhanced rather then suppressed by the small-mass effects.

\medskip

There also exist scattering processes for anti-fermions on $M_1$ monopoles that are obtained from~\eqref{eq:flipM1s}
by $CP$-conjugation,
%\begin{eqnarray}
%(\overline{u})_{R\,1,2}\, +\, M_1 \,&\to&\ (\overline{u})_{L\,1,2}\,+\, M_1 \nonumber\\
 %(\overline{d}\,)_{L\,3}\, +\, M_1 \,&\to&\ (\overline{d}\,)_{R\,3}\,+\, M_1 \nonumber \\
%(\overline{e})_{L}\, +\, M_1 \,&\to&\ (\overline{e})_{R}\,+\, M_1 
% \label{eq:flipM1sCP}
%\end{eqnarray}
%
\begin{align}
	(\psi)^\dagger_{\,I} +M\,\to\, (\tilde\psi)^\dagger_{\,I}+M\,: \quad %\left\{
	\begin{aligned}
\,\,(\overline{u})_{R\,1,2}\, +\, M_1 \,&\to\, (\overline{u})_{L\,1,2}\,+\, M_1   \\
\,\,(\overline{d}\,)_{L\,3}\, +\, M_1 \,&\to\, (\overline{d}\,)_{R\,3}\,+\, M_1    \\
 (\overline{e})_{L}\, +\, M_1 \,&\to\, (\overline{e})_{R}\,+\, M_1 
\end{aligned}
% \right.
 \label{eq:flipM1sCP}
\end{align}
Under the action of $CP$ the SM fermions change to the corresponding anti-fermions; 
the monopole, on the other hand, does not change to its anti-particle, it is odd under $P$ and under $C$, and thus
$ CP:\,\, M \, \to M$.

\subsection{$(B+L)$-violating scattering on the monopole}

We now turn to $(B+L)$-violating reactions induced by the monopole. The arguably most famous process of this type is the Callan-Rubakov scattering~\cite{Rubakov:1981rg,Rubakov:1982fp,Callan:1982ah,Callan:1982au}, 
\begin{equation}
u_L^1 \,+\, u_L^2\, +M_1\, \,\to\,
(\overline{e})_{R} \,+\, (\overline{d}\,)_{R\,3} +M_1
\label{eq:RC2f}
\end{equation}
which enables the monopole catalysis of the proton decay,
\begin{equation}
p^+\, +M_1\, \,\to\,
e^+\,+\, M_1 \,+\, {\rm pions}
\label{eq:proton}
\end{equation}
The process~\eqref{eq:RC2f} is unsuppressed by large UV scales of the underlying microscopic theory, it is also not suppressed by any 
exponentially small semiclassical factors~\cite{Rubakov:1988aq}.
The parton-level reaction~\eqref{eq:RC2f} is a $3 \to 3$ process where two 
$u$ quarks scatter on the minimal GUT monopole in the $j_0$-wave and produce the final state that changes $B+L$ by $-2$ units. 
The QED electric charge and the sum of $q_J$'s are conserved, and so is the flavour symmetry. The latter statement means that the initial and final state fermions transform in the same way under the $SU(3)_c$ in the sense that the combination of two $\boldsymbol3$ representations, $\epsilon_{ijk} \psi^j \psi^k$, transforms in the same way as the $\bar\psi_i$ in $\boldsymbol{\bar 3}$. 
Rubakov demonstrated in~\cite{Rubakov:1981rg,Rubakov:1982fp,Rubakov:1988aq} that the process~\eqref{eq:proton} can be generated by an instanton configuration in the monopole background in the UV theory, it can be interpreted as the effect of the chiral anomaly and is unsuppressed neither by the monopole mass nor exponentially.

The principal question then is what is the more elementary $(B+L)$-violating process involving a single fermion scattering on the monopole in the initial state.
Callan have argued in~\cite{Callan:1982ah,Callan:1982au,Callan:1982ac}  
that such process
must involve outgoing states with fractional fermion numbers (the so-called semitons or half-solitons arising in the bosonization picture of the $j_0$-wave fermions) and are of the form,
\begin{eqnarray}
 u_L^{\,1}\, +\, M_1 \,&\to&\, \frac{1}{2}\left(u_R^{1}\, (\overline{u})_{L2}\,(\overline{d}\,)_{R3} \,(\overline{e})_{R}\right)
 \,+\, M_1 \nonumber\\
d_R^{\,3}\, +\, M_1 \,&\to&\,  \frac{1}{2}\left(d_L^{\,3} \,(\overline{u})_{L1}\, (\overline{u})_{L2}\,(\overline{e})_{R}\right)
\,+\, M_1 \label{eq:callan}\\
 e_R\, +\, M_1 &\to&  \frac{1}{2}\left(e_L\,(\overline{u})_{L1}\, (\overline{u})_{L2}\,(\overline{d}\,)_{R3} \right)
 \,+\, M_1 \nonumber
\end{eqnarray}
Outgoing states in~\eqref{eq:callan} consist of four semitons, each with a fermion number 1/2. 
These states exist only in the massless theory and are destabilised when mass effects due to EWSB or the effects of strong interactions -- quark condensates and confinement --  become important.
The conventional interpretation~\cite{Callan:1982au,Callan:1983tm,Sen:1983yq,Preskill:1984gd}, 
see also~\cite{Dawson:1983cm,Kitano:2021pwt,Hamada:2022eiv},
is that at distances from the monopole greater than the inverse fermion mass/confinement/EWSB scale the semitons 
cease being good intermediate states, they somewhat miraculously disintegrate into states with integer fermion numbers that live in the usual perturbative Fock space.
In essence, like in the parton model, there is only a probability of finding a fermion of a given flavour in the final state at the cross-section level. 
In this analogy semitons play the role of  partons and quarks and leptons emerge as (stable) hadrons.
The probability of finding either of the four SM fermions in the final state in~\eqref{eq:callan} is of order one~\cite{Callan:1983tm,Sen:1983yq,Preskill:1984gd}.

\medskip

Recent progress by van Beest et al in~\cite{vanBeest:2023dbu,vanBeest:2023mbs}
provides a sharper understanding of the final states Fock space in the massless limit of the theory. In their construction the 
the outgoing radiation carries integer fermion numbers, but it is connected to the monopole by a topological surface. 
The existence of this surface provides an explicit realisation of the entanglement between the magnetically and electrically charged particles in the outgoing state. Being topological, the surface carries no energy but adds a twist operator to the final state. This ensures that the Fock states of the outgoing fermions are in the so-called twisted sector and are distinct from the original perturbative Fock space
 of incoming fermions. The resulting elementary $(B+L)$-violating processes for a SM fermion scattering on a monopole $M_1$
 are~\cite{vanBeest:2023mbs},
 \begin{equation}
 \psi^{\,I}+ M\,\to\, \tilde\psi^{\,I}+T+M\,
 \label{eq:Tproc1}
 \end{equation}
This is the lowest-partial-wave scattering process where both incoming $\psi^{\,I}$ and the outgoing $\tilde\psi^{\,I}$ fermions are in the $j_0$-wave and are described by 
by wave functions in a two-dimensional theory on $(r,t)$ half-plane. The twist $T$ indicates the presence of the topological surface stretched between the monopole and the outgoing fermion $\tilde\psi^{\,I}$. In the dimensionally-reduced theory the surface is a topological line from the
boundary at $r=0$ to the right-moving end-point at the position of the outgoing fermion $\tilde\psi^{\,I} (x)$. Using the state--operator correspondence in a two-dimensional theory, $T$ can be described by a local operator $T(x)$, at the same space-time point as the outgoing fermion, and trailed by the topological line from $x$ back to the monopole position at the origin. In Euclidean space this line represents a branch cut~\cite{vanBeest:2023mbs}: it picks up a non-trivial monodromy for any of the outgoing fermions $\tilde\psi^{\,J} (x)$, $J=1,\dots,N_f$, crossing it, so that,
\begin{align}
	\begin{aligned}
\tilde \psi^{\,J} (x) \, &\to&\,  e^{-2\pi i} \, \tilde \psi^{\,J} (x)   \\
\tilde \psi^{\,I\neq J} (x) \, &\to& 1\,\cdot\, \tilde \psi^{\,I\neq J} (x)
\end{aligned}
\quad : \quad T (x) \, \to\, \, e^{2\pi i /(N_f/2)} \,T (x)
 \label{eq:Tmonodr}
\end{align}
and no monodromy for the incoming fermions.
Equation~\eqref{eq:Tmonodr} implies that one can construct a flavour-singlet local twist operator that is invariant under all unbroken symmetries of the model and has the same quantum numbers as
$\left( (\tilde \psi(x))^\dagger_{\,1}\ldots (\tilde \psi(x))^\dagger_{\,N_f} \right)^{2/N_f}$.

For the minimal monopole case, the incoming fermions $\psi^{\,I}$ and the outgoing $\tilde\psi^{\,I}$ are the ones listed in Table~\ref{Tab:M1ferm}, where $I=1,\dots,N_f=4$, and the 
 operator $T$ is defined as the square root of the product over all flavours of the anti-fermions $(\tilde\psi)^\dagger$, which we denote as
 \begin{equation}
 T \,:=\,  \frac{1}{2}\left(\prod_{I=1}^{4} (\tilde\psi)^\dagger_I\right)\,=\, 
 \frac{1}{2}\left((\overline{u})_{L1}\, (\overline{u})_{L2}\,(\overline{d}\,)_{R3} \,(\overline{e})_{R}\right)
 \label{eq:T1def}
 \end{equation}
 The twist operator carries quantum numbers of the vacuum of the Fock space of outgoing fermions. These quantum numbers are consistent with all unbroken symmetries of the model: $T$ is a flavour singlet, it has a vanishing electric charge, $B-L=0$ and $B+L=1$. 
 Equation~\eqref{eq:T1def} implies that 
 it has the quantum numbers of a composite 
 anti-proton--positron state,
  \begin{equation}
 T(x)  \, =\,  \frac{1}{2}\Bigl(\bar p(x) \,\bar e(x)\Bigr)
 \end{equation}
 One should not be overly concerned about a superposition of states with different $B$ and $L$ charges since they are not good quantum numbers due to the anomalous nature of $B+L$ `symmetry'.
 
 \medskip
 
Taking into account the fact that fermion-anti-fermion pairs with same quantum numbers can annihilate, for example $(\overline{e})_{L} e_L = 1 \, + {\rm radiation}$,
it is easy to see that in practice this picture 
%\begin{align}
%	\psi^{\,I} +M_1\,\to\, \tilde\psi^{\,I}+T+M_1: \quad %\left\{
%	\begin{aligned}
%\,&u_L^{\,1,2} +M_1 \,\to\, u_R^{\,1,2}+ \frac{1}{2}\left((\overline{u})_{R1}\, (\overline{u})_{R2}\,(\overline{d}\,)_{L3} \,(\overline{e})_{L}\right)+ M_1   \\
%\,&d_R^{\,3} +M_1 \,\to\, d_L^{\,3}+ \frac{1}{2}\left((\overline{u})_{R1}\, (\overline{u})_{R2}\,(\overline{d}\,)_{L3} \,(\overline{e})_{L}\right)+M_1     \\
%\, &e_R + M_1 \,\to e_L+\frac{1}{2}\left((\overline{u})_{R1}\, (\overline{u})_{R2}\,(\overline{d}\,)_{L3} \,(\overline{e})_{L}\right)+ M_1   
%\end{aligned}
% \right.
 %\label{eq:flipM1T}
%\end{align}
\begin{align}
	\psi^{\,I} \,+\, M_1\,\to\, \tilde\psi^{\,I}+T\,+\,M_1: \quad %\left\{
	\begin{aligned}
\,&u_L^{\,1,2} +M_1 \,\to\, u_R^{\,1,2}+ T+ M_1   \\
\,&d_R^{\,3} +M_1 \,\to\, d_L^{\,3}+ T\,+\,M_1     \\
\, &e_R + M_1 \,\to e_L+T\,+\, M_1   
\end{aligned}
% \right.
 \label{eq:flipM1T}
\end{align}
provides an equivalent realisation of Callan's semiton processes~\eqref{eq:callan}. Note the processes~\eqref{eq:Tproc1} appear to be structurally very similar to the $(B+L)$-conserving reactions~\eqref{eq:flipM1s}.
The only difference between the them is the presence or absence of the twist operator on the right hand side characterising the Fock state of out going fermions. 

In the strictly massless limit of the (1+1)-dimensionally reduced theory on $(r,t)$ half-plane,
 the outgoing fermion state  
$\tilde\psi^{\,I}+T$ should be interpreted as a single fermion state~\cite{vanBeest:2023dbu,vanBeest:2023mbs}.
 The components of $\tilde\psi^{\,I}(x)$ and $T(x)$ all move together to the right with the speed of light and individual fermion components do not separate. The state $\tilde\psi^{\,I}+T$ represents the fermion attached to the topological defect that carries no energy or momentum. An experimentalist would observe a state that has the meaning of a single fermion in the modified (or twisted) Hilbert space. In terms of the original Hilbert space where the incoming fermions were defined, this is a complicated superposition, though it carries the same electric charge (and all other charges of non-anomalous symmetries of the theory) as the incoming fermion.
In any realistic scenario that allows for a particle interpretation, we need to include effects of fermion masses and that would allow to discuss individual fermion components.

The effects of fermion masses $m$ become significant at distances  $r \gtrsim 1/m$ where the surface connecting the monopole to the outgoing fermion
$ \tilde\psi^{\,I}$ becomes non-topological and breaks down. 
The intuitive picture  in many ways parallels the disintegration of semitons in the Callan approach. The main difference is that in the twisted sector approach, the topological surface did not carry any  energy (at least while stretched over lengths not too far from the monopole centre where mass effects have not set in yet), while fractionally charged semitons did.
What we call asymptotic final states are the states  defined at large distances $r \gtrsim 1/m$ and, with no topological surface left and no twist operator, their Fock space 
is in the conventional perturbative Fock space of outgoing massive fermions of definite fermion numbers.
Like in the Callan's original semiton picture, the outgoing fermion state $\tilde \psi^I +T$ in a massive theory becomes a superposition 
of fermion states with fractional fermion numbers. 
The latter determine the probability of finding a fermion of a given flavour in the final state at the cross-section level, as was originally suggested in Refs.~\cite{Callan:1983tm,Sen:1983yq,Preskill:1984gd}.

\medskip

We would like to conclude this section with a note that both: the semiton picture by Callan and the twisted Fock space sector
 approach of~\cite{vanBeest:2023dbu}, are based on the analysis of boundary conditions for fermions at the monopole centre. For the lowest partial wave this corresponds to matching incoming to outgoing fermions at $r=0$ in a 2-dimensional theory on half plane in $(r,t)$ coordinates. 
How does this relate to scattering amplitudes and what form to they take analytically?  The amplitudes formalism using conventional helicity spinor basis for the individual particles and the pairwise helicities for pairs of electric and magnetic states, was developed and applied in~\cite{Csaki:2020inw} to processes involving conventional states with integer particle numbers~\eqref{eq:flipM1s}. Subsequently scattering amplitudes for processes with fractional fermion numbers were constructed in~\cite{Khoze:2023kiu}.
The existence of analytic expressions for such amplitudes respecting all symmetries of the theory provides an additional justification of the validity of the scattering reactions~\eqref{eq:callan}, \eqref{eq:flipM1T}.
Amplitude expressions in~\cite{Khoze:2023kiu} contain fractional $2/N_f$ powers of spinor products for fractionally charged states. These result in branch cuts
that can be interpreted as arising from $\tilde \psi$-type fermions crossing the topological string extended from the monopole in the approach of~\cite{vanBeest:2023dbu}.

%%%%%%%%%%%%%%%%%%%%%%%%
\section{Phenomenological interpretation and cross-sections}
\label{Sec:inter1}
%%%%%%%%%%%%%%%%%%%%%%%%

We now need to understand what the formal expressions for $j_0$-wave scattering on the monopole, 
tell us about scattering of physical SM states and derive cross-sections that are measurable quantities.

We have learned that there are three stages of intermediate would-be-asymptotic final states. Gradually increasing the distance $r$ from the monopole, we first have semitons or the twisted-sector states, then SM partons, and finally colour-neutral states - hadrons and leptons.

%%%%%%%%%%%%%%%%%%%%%%%%
\subsection{Processes initiated by a lepton}
%%%%%%%%%%%%%%%%%%%%%%%%

In discussing any scattering process we need to start with an incoming asymptotic state. We first consider the $e_R$-initiated processes -- the third process
in~\eqref{eq:callan} or~\eqref{eq:flipM1T}.
The incoming electron $e_R$ and the monopole $M$ are both good asymptotic state and we can perform the partial-wave decomposition of the initial state
and consider independently the scattering in the lowest available partial-wave $j=j_0$ and the scattering in all higher waves,
\begin{eqnarray}
j=j_0:\qquad
&e_R \,+\, M \,\to\, e_L\,+\,T\,+\, M 
\label{eq:eM1j0} \\
j \ge j_0+1 :\qquad
&e_R \,+\, M \,\to\, e_R\,+\, M 
\label{eq:eM1jh}
\end{eqnarray}
The second process~\eqref{eq:eM1jh} is
the already well-understood KYG helicity non-flip scattering of the type \eqref{eq:hnfKYG} and its cross-section has been computed and tabulated
in~\cite{Kazama:1976fm}.

The first process~\eqref{eq:eM1j0} requires some care in disentangling the structure of the physical final states. We have already noted that the twist operator 
$T$ defined in~\eqref{eq:T1def}, has the structure of a composite anti-proton--positron state with the coefficient 1/2 which implies that it describes a linear combination of two different final states, or in other words, it describes two scattering outcomes:
\begin{eqnarray}
e_R \,+\, M &\to&\, e_L\,+\, M 
\label{eq:eM1j0ee} \\
e_R \,+\, M &\to&\, \overline{p}_L\,+\, M 
\label{eq:eM1j0ep}
\end{eqnarray}
We treat these as inclusive processes meaning that the final states can also include any additional radiation $X$, which conserves all quantum numbers including electric charge, $B$ and $L$. The first process conserves $B+L$ while the second has  $\Delta(B+L)=-2$.

We find that cross-sections for each~\eqref{eq:eM1j0ee} and~\eqref{eq:eM1j0ep} are then determined by the KYG helicity-flip expressions times the appropriates branching ratios. In the CoM frame we have,
\begin{align}
	\frac{d \sigma}{d \Omega} \,=\, \frac{q_J^2}{(p_c^{\,e})^2}\, \left(\sin (\theta/2) \right)^{4|q_J|-2}\, \times
	\, \left\{
	\begin{aligned}
\,& {\rm br}_{e}=\, 1- \frac{1}{2}(p_c^{\,\bar p}/p_c^{\,e})
\qquad{\rm for}\quad  e_R + M \,\to\, e_L+ M      \\
\, & {\rm br}_{\bar{p}}=\,\frac{1}{2}(p_c^{\,\bar p}/p_c^{\,e})
\,\,\,\,\quad \qquad{\rm for}\quad  e_R + M \,\to\, \overline{p}_L+ M 
\end{aligned}
\right.
\label{eq:flipsig-e}
\end{align}
where $p_c^{\,e}$ and $p_c^{\,\bar p}$ denote the CoM momenta of the electron and the anti-proton and
$\theta$ is the polar angle of the scattered monopole in the CoM frame. For future reference, we have kept 
$q_J$ general. For the minimal monopole $M_1$ the pairwise helicity of the electron and the anti-proton is $q_J=-1/2$, the dependence on $\sin(\theta/2)$ disappears 
and~\eqref{eq:flipsig-e} takes the simple form %$\frac{d \sigma}{d \Omega} \,=\, \frac{1}{(2 p_c^{\,e})^2}\, {\rm br}_{e/\bar p}$.
${d \sigma}/{d \Omega} \,=\, 1/{(2 p_c^{\,e})^2}$ times the relevant branching ratio ${\rm br}_{e/\bar p}$.

The branching ratio ${\rm br}_{\bar{p}}$ for the anti-proton in the final state used in the second equation in~\eqref{eq:flipsig-e} was obtained  from the ratio of the 2-particle phase space
volumes in the final state,
\begin{equation}
{\rm br}_{\bar{p} }\,=\, \frac{1}{2} \, {\rm Vol}\, \Phi_{\bar{p} M} /  {\rm Vol}\, \Phi_{e M}
\end{equation}
When the kinetic energy carried by the electron becomes insufficient to produce an on-shell anti-proton, $p_c^{\,\bar p}\to 0$, which implies that ${\rm br}_{\bar{p} }\to 0$ and $ {\rm br}_{e}\to 1$.
In the opposite high energy limit each of the branching ratios approach 1/2. 

The dependence of the overall rate in~\eqref{eq:flipsig-e} on the initial-state electron momentum implies that for non-relativistic collisions 
the cross-section of the process~\eqref{eq:eM1j0ee},
\begin{equation}
\frac{d \sigma_{e_R\to e_L}}{d \Omega} \,=\, \frac{q_J^2}{(p_c^{\,e})^2}\, \left(\sin \theta/2 \right)^{4|q_J|-2}
\,\,\sim \,\,\frac{q_J^2}{(m_e)^2} \, \frac{1}{\beta^2}
\label{eq:eM1j0eesig} 
\end{equation}
is very large, being enhanced both by the smallness of the electron mass and by the non-relativistic electron's velocity $\beta = v_e/c$.

\medskip

The anti-proton synthesis process~\eqref{eq:eM1j0ep} occurs only when the electron's kinetic energy is sufficient to produce an on-shell proton,
$p_c^{\,e} > m_p$, and
\begin{equation}
\frac{d \sigma_{e \to \bar{p}}}{d \Omega} \,\,\sim \, \, q_J^2\, \frac{p_c^{\,\bar{p}}}{(p_c^{\,e})^3}\, < \,q_J^2\,
\frac{1}{({\rm GeV})^2} 
\end{equation}

\medskip

We do not need to consider separately the $(B+L)$-conserving process (last equation in~\eqref{eq:flipM1s}) discussed in section~\ref{sec:5.1} as it is 
exactly the same as the process~\eqref{eq:eM1j0ee} 
investigated here. 
It is interesting to note that its cross-section~\eqref{eq:eM1j0eesig} is large,  which is opposite to what one might have naively expected based on the $SU(2)_L$ gauge-invariance arguments reviewed in section~\ref{sec:5.1}. The reason why this is the case is the EWSB, as we have anticipated in section~\ref{sec:5.1}. 
Such $(B+L)$-conserving  processes turn out to be  enhanced rather than suppressed by small fermion masses $m$ in the non-relativistic limit.
\medskip

%%%%%%%%%%%%%%%%%%%%%%%%
\subsection{Processes initiated by quarks}
%%%%%%%%%%%%%%%%%%%%%%%%

Our next goal is to estimate cross-sections for $(B+L)$-violating processes initiated by quarks~\eqref{eq:flipM1T}.

A key ingredient for studying scattering of fermions on monopoles is the partial-wave decomposition and at the outset it needs to be applied to proper 
asymptotic incoming states. 
For quark-initiated microscopic processes in~\eqref{eq:flipM1T} the incoming states are not asymptotic at distances greater than the QCD confinement 
scale $1/\Lambda_{QCD} \sim 1{\rm fm}$ so at least initially we have to start with a proton. Since the infinitely separated proton and the monopole form a good asymptotic incoming state, we proceed to expand it in partial-waves $j$ and then, in analogy with our approach in the previous section, we select the 
lowest $j_0$-wave.
For a non-relativistic proton we should find a combination of two KYG-type helicity flip reactions, in direct analogy with what we have done 
in~\eqref{eq:eM1j0ee}-\eqref{eq:eM1j0ep},
\begin{eqnarray}
p_L \,+\, M &\to&\, p_R\,+\, M 
\label{eq:eM1j0pp} \\
p_L \,+\, M &\to&\, \overline{e}_R\,+\, M 
\label{eq:eM1j0ppee}
\end{eqnarray}
These are treated as inclusive processes; the first process conserves $B+L$ and the second violates it by $\Delta(B+L)=-2$. 
The proton for now is considered to be sufficiently non-relativistic, so that its internal partonic structure does not play a role. 
In this case, the proton acts as an elementary particle and
can fall as a whole on the monopole centre. For this to be correct, we need 
the proton momentum to be less than the inverse size of the proton, so that the individual partons do not escape the pull of the strong interaction force and the proton can fall on the monopole core,
\begin{equation}
p_c^{\,p} \,=\, m_p \, \beta\,  \gamma \, <  1\,{\rm fm}^{-1} \quad \Rightarrow \quad \beta < 0.2
\label{eq:NRbeta}
\end{equation} 
For the cross-sections of these processes in the CoM frame we then have,
\begin{align}
	\frac{d \sigma}{d \Omega} \,=\, \frac{q_J^2}{(p_c^{\,p})^2}\, \left(\sin (\theta/2) \right)^{4|q_J|-2}\, \times
	\, \left\{
	\begin{aligned}
\,& {\rm br}_{p}=\, \frac{p_c^{\,p}}{ p_c^{\,\bar e}+p_c^{\,p}}
\,\,\,\,\quad{\rm for}\quad  
                 p_L + M \to\, p_R+M     \\
\, & {\rm br}_{\bar{e}}=\,\frac{p_c^{\,\bar e}}{ p_c^{\,\bar e}+p_c^{\,p}}
\,\,\,\, \quad{\rm for}\quad  
               p_L + M \to\, \overline{e}_R + M 
\end{aligned}
\right.
\label{eq:flipsig-p}
\end{align}
In summary, we conclude that for a non-relativistic proton with  $\beta < 0.2$ our estimate for the cross-section
for the monopole catalysis of the proton decay process~\eqref{eq:eM1j0ppee} is
\begin{equation}
\frac{d \sigma_{\, p\to \,\bar e+X}}{d \Omega} \,\simeq\, \frac{q_J^2}{\beta^2}\,
\frac{1}{m_p^2}\,\, \left(\sin \theta/2 \right)^{4|q_J|-2}
%\,\sim \,\frac{1}{(m_p)^2} \, \frac{1}{\beta^2}
\quad{\rm for}\quad  \beta < 0.2
\label{eq:eM1pdecay} 
\end{equation}
which is consistent with~\cite{Rubakov:1983sy,Rubakov:1988aq}.

\medskip

We now how the above analysis is modified for a relativistic incoming proton.
In this case we can still select the initial proton--monopole state to be in the $j_0$ wave. 
When such a proton reaches the monopole, the energetic valence quarks are spread over the area of the proton radius.
With their momenta being $\sim p_c^p \gg \Lambda_{QCD}$, they pass the monopole far away from its core. To ensure that they can hit the monopole core we have to select the $u$ and $d$ quark states with minimal impact parameters -- this is equivalent to re-selecting $j_0$-waves now for sea quarks inside the proton, so that the quark-initiated processes  
\begin{eqnarray}
 u_L^{\,1}\, +\, M_1 \,&\to&\, \frac{1}{2}\left(u_R^{1}\, (\overline{u})_{L2}\,(\overline{d}\,)_{R3} \,(\overline{e})_{R}\right)
 \,+\, M_1 \nonumber\\
  u_L^{\,2}\, +\, M_1 \,&\to&\, \frac{1}{2}\left(u_R^{2}\, (\overline{u})_{L1}\,(\overline{d}\,)_{R3} \,(\overline{e})_{R}\right)
 \,+\, M_1
 \end{eqnarray}
or their combination,
\begin{equation}
u_L^1 \,+\, u_L^2\, +M_1\, \,\to\,
(\overline{d}\,)_{R\,3} +(\overline{e})_{R} \,+\, M_1
\label{eq:RC2fRub}
\end{equation}
can take place.
This costs the geometric suppression factor of \newline
$\xi \,=\, (R_M/R_p)^2 \sim (m_p/M_{M})^2 \ll 1$ in the KYG cross-section.
The parameter $\xi$ plays the role of the inelasticity parameter which was found in Refs.~\cite{Mohapatra:1997sc,Chung:1997rz}
to be similarly tiny for collisions of protons with some hypothetical very massive colour-singlet particles. Our geometric discussion here followed the same idea.

As the result, the $j_0$-wave proton-monopole 
CoM cross-sections in~\eqref{eq:flipsig-p}  should be sharply cut-off at $p_c^p > \Lambda_{QCD}$. 
For relativistic protons we have essentially recovered the usual $ 1/M_M^2\sim 1/s$ suppression 
characteristic to perturbative $s$-channel processes, which slower protons with $p_c^p < \Lambda_{QCD}$ have been able to avoid.

%%%%%%%%%%%%%%%%%%%%%%%%
\section{Electroweak monopoles in the Standard Model}
\label{Sec:ewM}
%%%%%%%%%%%%%%%%%%%%%%%%

The triple-charged monopoles $M_3$ and  $M_3^\prime$ are special in the Standard Model -- they do not have chromomagnetic flux. Unlike the minimal GUT monopole $M_1$ there is no ambiguity with how to modify their classical  field configurations at $r > 1{\rm fm}$ to account for confinement; $M_3$ and  $M_3^\prime$ have no $SU(3)_c$ gauge field components.

%%%
\subsection{Electroweak $M_3$ monopole with no EWSB}
\label{sec:M3}
%%%

 \begin{table}[ht]
\begin{equation} \nonumber
\begin{array}{lc|cccc}
\quad {M_3} \text{ monopole}
&&\quad q_J &\,SU(2)_{J} & \,SU(3)_c\, & \,Q_{el}\,
\\\hline
&u_L^i\,\,&+\frac{1}{2}&\boldsymbol1&\boldsymbol3&+\frac{2}{3}
\\
\psi\,\, \text{incoming:}  
&d_R^{\,i}\, & -\frac{1}{2}&\boldsymbol1& {\boldsymbol3}& -\frac{1}{3}
\\
&\ e_R\,& -\frac{3}{2} &\boldsymbol3 &\boldsymbol1&-1
 \\ \hline 
&u_R^i &+1 &\boldsymbol2& \boldsymbol3& +\frac{2}{3}
\\
\tilde\psi\,\, \text{outgoing:}  
&e_L &-1 &\boldsymbol2& \boldsymbol1& -1
\\
&(\overline{\nu})_R&-\frac{1}{2}& \boldsymbol1& \boldsymbol1& 0
\end{array}
\end{equation}
\caption{SM fermions and their charges in the $j_0$ partial-wave scattering on the electroweak $M_3$ monopole. All three colours $i=1,2,3$ are interacting 
for $u_L$, $d_R$ and $u_R$, however $d_L$ quarks have vanishing $q_J$ and are missing.}
\label{Tab:M3ferm}
\end{table}

We first consider the electroweak monopole $M_3$ which can exist in the SM before EWSB. Its magnetic fluxes are listed in the row in Table~\ref{Tab:Mpclass}.  We follow the same strategy as in the beginning of section~\ref{Sec:minM} and proceed to compute the pairwise helicities $q_J$ of $M_3$ with each of the SM fermions using~\eqref{eq:qJSMgen} to find which of them have non-vanishing interactions with $M_3$. We list these fermions in Table~\ref{Tab:M3ferm} along with their charges and in/out state assignments determined as before by the 
 sign of $q_J$.  Importantly, these fermions live in non-trivial representations of the rotation group  $SU(2)_{J}$, which is different from what was the case
 for fermions interacting with  $M_1$.

Based on these charges and representations assignments, the authors of~\cite{vanBeest:2023mbs} have noticed that is impossible to
make a meaningful map between $\psi^I$ and $\tilde \psi^I$ fermions to enable scattering processes of the type
 \begin{equation}
 \psi^{\,I}+ M\,\to\, \tilde\psi^{\,I}+T+M\,
 \label{eq:Tproc3}
 \end{equation}
There are also problems with constructing $T$ from the product of all $N_f=9$ of the outgoing fermions 
$(\tilde\psi)^\dagger_{\,I}$ as this product carries a non-vanishing electric charge. 
It is not even obvious how to construct an instanton-dominated process in the monopole background as it would have to have $9/2$ incoming fermions going into $9/2$ outgoing ones.
While this apparent inability to construct a meaningful fermion scattering on the $M_3$ monopole seems worrying~\cite{vanBeest:2023mbs}, it does not actually apply to the Standard Model settings when the electroweak symmetry os broken by the SM Higgs vev. We also note that it is not a priori clear whether it makes sense to restrict external states scattering on $M_3$ in the purely massless theory solely to fermions. Participation of massless non-Abelian gauge fields that can carry electric charges in scattering processes 
could be important.

As we already noted earlier, the breaking of the electroweak symmetry in the Standard Model does affect non-trivially the triple-charged monopole and 
through interactions with the SM Higgs field modifies its magnetic $SU(2)_L$ flux. The modified stable monopole in the SM with EWSB is the 
electromagnetic monopole $M_3^{\,\prime}$ in Table~\ref{Tab:Mpcpr}.

%%%
\subsection{Minimal electromagnetic $M_3^{\,\prime}$ monopole in the SM}
\label{sec:M3prime}
%%%

The triple-charged monopole consistent with EWSB in the SM  is the purely electromagnetic $M_3^{\,\prime}$. It is the minimal monopole that has only 
QED electromagnetic long range interactions and is arguably the most likely monopole candidate surviving in the SM.  The fact that it does not carry
QCD chromomagnetic flux means that it neither confines nor needs to be otherwise modified by dynamical quantum effects of strong interactions. This
distinguishes it from the minimal GUT monopole $M_1$ which is not a colour-singlet. In fact, if  $M_1$ monopoles confine, they would form baryon-like bound states with attractive triple string junctions pulling them, together. This should result in a stable colourless configuration with minimal energy in charge-3 sector, which is the
$M_3^{\,\prime}$ monopole. 

Following the same approach as in sections~\ref{Sec:minM} and~\ref{sec:M3} we determine the charges and representations of SM fermions 
that can scatter on $M_3^{\,\prime}$. This information is assembled in Table~\ref{Tab:M3ferm-pr}.

\begin{table}[ht]
\begin{equation} \nonumber
\begin{array}{lc|cccc}
\quad {M_3^{\,\prime}} \text{ monopole} &&\quad q_J &\,SU(2)_{J} & \,SU(3)_c\, & \,Q_{el}\,
\\\hline
&\,\, u_L^{\,i,s}\,& +1& \boldsymbol2& \boldsymbol3&+\frac{2}{3}
\\
\psi^I\,\, \text{incoming:}  
&d_R^{\,i}\, &-\frac{1}{2}&\boldsymbol1&{\boldsymbol3}& -\frac{1}{3}
\\
&\ e_R^{\,a}\,& -\frac{3}{2} &\boldsymbol3 &\boldsymbol1&-1
 \\ \hline 
&\,\, u_R^{\,i,s}\,& +1& \boldsymbol2 & \boldsymbol3&+\frac{2}{3}
\\
\tilde\psi^I\,\, \text{outgoing:}  
&d_L^{\,i}\, &-\frac{1}{2}&\boldsymbol1&{\boldsymbol3}& -\frac{1}{3}
\\
&\ e_L^{\,a}\,& -\frac{3}{2} &\boldsymbol3&\boldsymbol1&-1
\end{array}
\end{equation}
\caption{SM fermions and their charges in the $j_0$ partial-wave scattering on the purely electromagnetic SM monopole $M_3^{\,\prime}$. There are $N_f=12$ flavours of $\psi^I$ and the same number of $\tilde \psi^I$.}
\label{Tab:M3ferm-pr}
\end{table}

Note that $u_L^{\,i,s}$-quarks form the doublet ($s=\pm1/2$) representation of the rotation group $SU(2)_J$, $d_R^{\,i}$-quarks are singlets and electrons 
$\ e_R^{\,a}$ transform in the triplet
 ($a=-1,0,1$) representation. The $SU(3)_c$ colour indices are $i=1,2,3$. There is a perfect symmetry between the incoming and outgoing fermion representations in Table~\ref{Tab:M3ferm-pr} and it is 
straightforward to pair $\psi^I$ and $\tilde\psi^I$ fermions for each $I=1,\ldots, N_f=12$ simply by flavour, i.e. in the order they appear in the Table.
Hence we can write down the single fermion scattering processes with the twist, following the same general logic as in~\eqref{eq:Tproc1},

\begin{align}
	\psi^{\,I} + M_3^{\,\prime}\,\to\, \tilde\psi^{\,I}+T+M_1: \quad %\left\{
	\begin{aligned}
	\, &e_R^{\,a} + M_3^{\,\prime} \,\to e_L^{\,a}+T\,+\, M_3^{\,\prime} \\
\,&u_L^{\,i,s} +M_3^{\,\prime} \,\to\, u_R^{\,i,s}+ T+ M_3^{\,\prime}   \\
\,&d_R^{\,i} +M_3^{\,\prime} \,\to\, d_L^{\,i}+ T\,+\,M_3^{\,\prime}     
\end{aligned}
% \right.
 \label{eq:TprocM3pr}
\end{align}
with the twist operator $T$ being the $2/N_f=1/6$ power of the product over all flavours of the anti-fermions $(\tilde\psi)^\dagger$, 
 \begin{equation}
 T \,:=\,  \frac{1}{6}\left(\prod_{I=1}^{12} (\tilde\psi)^\dagger_I\right)\,=\, 
 \frac{1}{6}\left(((\overline{u})_{L})^6\, ((\overline{d}\,)_{R})^3 \,((\overline{e})_{R})^3\right)
 \label{eq:T3prdef}
 \end{equation}
 Each product of fermions is over all different components of the colour and the $SU(2)_J$ indices.
 
 The twist operator~\eqref{eq:T3prdef} is a singlet under all unbroken symmetries of the model: $T$ is invariant under $SU(3)_c$ and $SU(2)_J$ transformations, it has vanishing electric charge, $B-L=0$ and $B+L=1$. It  also can be interpreted as a composite 
 anti-proton--positron state, $T= \frac{1}{2}(\bar p\, \bar e)$, which is analogous to what we have seen  in the case of $M_1$. 

\medskip

Fermion scattering in the triple monopole background has been studied previously in~\cite{Rubakov:1983sy} and also in~\cite{Schellekens:1984wv}
by searching for
$SU(5)_{GUT}$-instanton generated 
multi-fermion processes in the monopole background (also in the presence of $s$-quarks and $\mu$'s) and resulting in a 
10-to-10 particle scattering involving 9 incoming and 9 outgoing fermions scattering on this monopole.
Without $s$-quarks and muons of the second generation we would expect the instanton process to be a little simpler and involve only 6 incoming $\psi$ and 6 outgoing $\tilde\psi$ fermions.
These are considerably more complicated than the more elementary processes in~\eqref{eq:TprocM3pr} and would correspond to six-fold superpositions 
of the latter.
 
 \medskip
 
 Processes in~\eqref{eq:TprocM3pr} can now be used to obtain scattering cross-sections of SM fermions following the approach developed in section~\ref{Sec:inter1}.
Electron-initiated processes lead to
\begin{eqnarray}
e_R^{\,a} \,+\, M_3^{\,\prime} &\to&\, e_L^{\,a}\,+\, M_3^{\,\prime}
 \\
e_R^{\,a} \,+\, M_3^{\,\prime} &\to&\, \overline{p}_L^{\,a}\,+\, M_3^{\,\prime}
\end{eqnarray}
where the electron as well as the anti-proton states are in the $j_0=1$ partial wave, they transform in the triplet representation 
of the rotation group and are labeled by $a=-1,0,1$.
The cross-sections given by the same expressions as in~\eqref{eq:flipsig-e} now with $|q_J|=3$.

Processes initiated by quarks are recast in terms of colour-singlet asymptotic states as,
\begin{eqnarray}
p_L^{\,a} \,+\, M_3^{\,\prime} &\to&\, p_R^{\,a}\,+\, M_3^{\,\prime}
\\
p_L \,+\, M_3^{\,\prime} &\to&\, \overline{e}_R\,+\, M_3^{\,\prime} 
\end{eqnarray}
and their cross-sections estimates are given in~\eqref{eq:flipsig-p}-\eqref{eq:eM1pdecay} with $|q_J|=3$.

This concludes or discussion of the scattering on the purely electromagnetic SM monopole.

%%%%%%%%%%%%%%%%%%%%%%%%
\section{The rest of the SM monopoles after EWSB}
\label{Sec:otherM}
%%%%%%%%%%%%%%%%%%%%%%%%

Having discussed in detail fermion scattering on the single-charged and the triple-charged monopoles, we now consider fermion interactions
with the remaining monopoles $M_2^{\,\prime}$, $M_4^{\,\prime}$, $M_5^{\,\prime}$, and $M_6^{\,\prime}$ in the SM after EWSB.

%%%%%%%%%%%%%%%%%%%%%%%%
%\subsection{$M_2^{\,\prime}$ monopole}
%\label{Sec:M2prime}
%%%%%%%%%%%%%%%%%%%%%%%%

Following the same logic as  in the previous section, we compute values of $q_J$ and assemble charges and representations for each of the SM quarks and leptons in the background of the monopole $M_2^{\,\prime}$. These are shown in Table~\ref{Tab:M2pr-f}.

 \begin{table}[ht]
\begin{equation} \nonumber
\begin{array}{lc|cccc}
\quad {M_2^{\,\prime}} \text{ monopole} &&\quad q_J &\,SU(2)_{J} & \,SU(2)_c\, & \,Q_{el}\,
\\\hline
&\,\, u_L^{\,1,2}\,& +\frac{1}{2}& \boldsymbol1& \boldsymbol2&+\frac{2}{3}
\\
&\,u_L^{\,3}\,& +1& \boldsymbol2& \boldsymbol1&+\frac{2}{3}
\\
\psi\,\, \text{incoming:}  
&\,\,\,d_R^{\,1,2}\, &-\frac{1}{2}&\boldsymbol1&{\boldsymbol2}& -\frac{1}{3}
\\\,
&\ e_R\,& -1 &\boldsymbol2 &\boldsymbol1&-1
 \\ \hline 
&\,\, u_R^{\,1,2}\,& +\frac{1}{2}& \boldsymbol1& \boldsymbol2&+\frac{2}{3}
\\
&\,u_R^{\,3}\,& +1& \boldsymbol2& \boldsymbol1&+\frac{2}{3}
\\
\tilde\psi\,\, \text{outgoing:}
&\,\,\,d_L^{\,1,2}\, &-\frac{1}{2}&\boldsymbol1&{\boldsymbol2}& -\frac{1}{3}
\\
&\ e_L\,& -1 &\boldsymbol2 &\boldsymbol1&-1
\end{array}
\end{equation}
\caption{SM fermions interacting with the $M_2^{\,\prime}$ monopole. All representations are symmetric between the incoming $\psi$ and the outgoing 
$\tilde\psi$ fermions. There are $N_f=8$ flavours of each.}
\label{Tab:M2pr-f}
\end{table}
%\tilde\psi\,\, \text{outgoing:}

We note that all charges and representations are symmetric between the incoming and the outgoing 
$\tilde\psi$ fermions. There is a straightforward map between $\psi^I$ and $\tilde\psi^I$ for each flavour $I=1,\ldots, N_f=8$. 
This is  the same pattern as was established for the $M_1$ and $M_3^{\,\prime}$ monopoles ({\it cf.}~Tables~\ref{Tab:M1ferm} and \ref{Tab:M3ferm-pr}).

\bigskip
 
 \noindent Scattering processes for the remaining  $M_4^{\,\prime}$, $M_5^{\,\prime}$, and $M_6^{\,\prime}$ monopoles retain a very similar pattern
 and  these monopoles do not need to be considered separately.
 We conclude that there are no obstacles in describing fermion scattering on any of the SM monopoles after EWSB.

%%%%%%%%%%%%%%%%%%%%%%%
\section{Summary}
\label{Sec:concl}
%%%%%%%%%%%%%%%%%%%%%%%

Our main motivation was two-fold. On the one hand we wanted to clarify to what extend the recently proposed in~\cite{vanBeest:2023dbu}
resolution of the 
paradox related to fractional particle numbers appearing in scattering processes of massless fermions with monopoles applies to the Standard Model.
On the other hand we also wanted to extract some practical tangible conclusions of what this means for physical processes and to estimate their
cross-sections. These involve processes with and without $(B+L)$-violation and the monopole-induced proton decay~\cite{Rubakov:1981rg,Rubakov:1982fp,Callan:1982ah,Callan:1982au}.

Since our primary goal was to deal with scattering processes in the Standard Model with all matter fields being massive,
we first needed to distinguish between monopole states in a massless non-Abelian theory and the monopoles that survive in the SM theory after the electroweak symmetry breaking. With the exception of the minimal GUT monopole, all these monopoles are different from those of the massless SM simply because their magnetic fluxes under the weak sector $SU(2)$ are different. 

We pointed out that for all monopoles descending from a completely generic UV theory, those which can survive in the present-day electroweak Higgs phase of the Standard Model pose no obstacles to have their scattering with the SM fermions described in the approach of~\cite{vanBeest:2023dbu}. 
This is different from the conclusions reached in~\cite{vanBeest:2023mbs} that the scattering problem cannot be currently solved for certain monopoles,
including $M_3$ -- the minimal monopole configuration with no chromomagnetic charge. We found that this problem is circumvented for all monopole configurations after EWSB takes place. The underlying reason for this is the fact that the SM monopoles modified by the effect of EWSB interact with 
the fermions that can be assembled into vector-like rather than chiral representations.

For all such monopoles,  $M_{1}',\ldots,M_{6}'$, scattering processes with a single fermion in the initial state are characterised 
are of the form,  
\begin{equation}
 \psi^{\,I}+ M\,\to\, \tilde\psi^{\,I}+T+M\,,  \qquad
 T \,=\,  \frac{2}{N_f}\left(\prod_{I=1}^{N_f} (\tilde\psi)^\dagger_I\right)
 \label{eq:Tproc-gen1}
 \end{equation}
where the twist operator ~\cite{vanBeest:2023dbu} is defined in terms of the product over all flavours of the outgoing anti-fermions $(\tilde\psi)^\dagger$.
 
 We showed that for the colour-singlet asymptotic states these result in physical Standard Model (inclusive) scattering processes with and without $(B+L)$-violation, proton synthesis and proton decay,
 \begin{eqnarray}
e_R \,+\, M \,&\to&\, e_L \,+\, M   \,\,\, \,+\, X
 \\
e_R \,+\, M  \,&\to&\,  \overline{p}_L\,+\, M   \,\,\, \,+\, X
\label{eq:9.3}
\\
p_L \,+\, M  \,&\to&\,  p_R\,+\, M   \,\,\, \,+\, X
\\
p_L \,+\, M  \,&\to&\,  \overline{e}_R\,+\,M  \,\,\, \,+\, X
\label{eq:9.5}
\end{eqnarray}
Cross-sections formulae for these processes are given in Eqs.~\eqref{eq:flipsig-e}, \eqref{eq:flipsig-p}-\eqref{eq:eM1pdecay} and we have discussed their numerical estimates.
The values of $|q_J|$ to be used in these expressions are the ones set by the electron (it is the same for the anti-proton) which for each monopole type are given by one half of the monopole topological charge $p$,
\begin{equation}
q_J (e^-, M_p^{\,\prime}) \,=\, - p/2 
\end{equation}

\bigskip

There are still many open questions and issues to be better understood and clarified in the physics of monopole scattering and interactions, even in the absence of any signs of monopoles being around, at least for now. 
For example, it is currently unknown how to identify an explicit QFT mechanism or a semi-classical configuration that generates  {\it minimal}
scattering processes~\eqref{eq:Tproc-gen1} on a monopole that involve a single fermion in the initial state in a theory with $N_f>2$.
All presently known instanton-induced processes in the monopole background necessarily involve all $N_f$ flavours of massless (or very light) fermions: $N_f/2$ in the incoming, and $N_f/2$ in the outgoing state~\cite{Rubakov:1982fp,Rubakov:1988aq}. 
These processes, such as~\eqref{eq:RC2fRub} 
for the minimal monopole, contribute to the catalysis of the proton decay, but not to the minimal process~\eqref{eq:flipM1T} at partonic level.
In general, the ability to identify a specific non-perturbative configuration or a mechanism depends on whether or not the theory allows for a semi-classical treatment of these processes. For example, the non-perturbative  Affleck-Dine-Seiberg (ADS) superpotential~\cite{Affleck:1984xz} in a supersymmetric QCD with $N$ colours and $N_f$ flavours is fully determined by the action of exact and anomalous symmetries of the theory, much alike the process~\eqref{eq:Tproc-gen1}.
But only in the special case of $N_f=N-1$ the ADS superpotential is generated by instantons. For $0\le N_f < N-1$ of massless flavours, the low-energy theory is a strongly-coupled $SU(N-N_f)$ gauge theory and instantons do not contribute to the superpotential. One can modify the the original theory by compactifying $R^4$ to $R^3 \times S^1$, and in this case it can be shown that the ADS superpotential is generated by fractional instantons, identified as monopole-like configurations in a 3-dimensional theory  for a small radius of $S^1$ where the theory is weakly coupled~\cite{Davies:1999uh}.  Very similar considerations also apply to the generation of gluino condensate in pure supersymmetric Yang-Mills theory (SYM)~\cite{Davies:1999uw}.
SYM on $R^4$ does not admit a semiclassical weak coupling approximation and a straightforward counting of instanton fermion zero modes implies that instantons of this theory do not contribute to the gluino condensate.
 On the other hand, if the the model is partially compactified on $R^3 \times S^1$, semiclassical instanton-monopole configurations do in fact generate the correct value for the gluino condensate in the path integral~\cite{Davies:1999uw}.

It is currently an open problem of whether it is possible to identify an appropriate semiclassical configuration dominating path integral (in an appropriately defined theory) that would generate minimal processes of the type~\eqref{eq:Tproc-gen1} that underpin proton catalysis.
A viable alternative, in our opinion, is to resort to Callan's original approach~\cite{Callan:1982au} where the minimal process~\eqref{eq:Tproc-gen1} is sourced in QFT by solitons carrying fractional fermion numbers, using the bosonization formalism in of the dimensionally reduced theory.

\medskip 

An interesting consequence of our results is that in so far as one can always rely on the minimal processes~\eqref{eq:Tproc-gen1} to be present,
the unsuppressed $(B+L)$-violating proton catalysis and proton decay processes~\eqref{eq:9.3},\eqref{eq:9.5} exist in all extensions of the SM that support any of the monopoles $M_p'$. At first sight, this result appears to contradict the prevailing wisdom in the early literature that the monopole catalysis processes are model-dependent and do not occur for for certain monopoles in certain non-minimal GUT extensions 
of the SM~\cite{Dawson:1982sc,Schellekens:1984wv,Rubakov:1988aq}. It should be noted, however, that the monopoles considered in~\cite{Dawson:1982sc,Schellekens:1984wv} were formed at an {\it intermediate stage}
of the $SO(10)_{GUT}$ breaking to $SU(4)\times SU(2)_L\times SU(2)_R$ with the unbroken left-right symmetry, thus containing long-range massless gauge field components along these subgroups.
 These are not the SM monopoles $M_p'$ considered in this paper as their gauge components are altered by the electroweak symmetry breaking, thus changing the pattern of fermion--monopole interactions. It should also be noted that most of the  discussion in the early literature concentrated on fermion multiplets involved in the {\it instanton-generated} processes, which, as we have noted earlier, involve a `maximal' number $N_f/2 \to N_f/2$ 
 of scattered fermions, and in general, do not constrain the more elementary minimal processes of the type~\eqref{eq:Tproc-gen1}. It would be interesting to investigate this further.

\medskip

\section*{Acknowledgements}

I thank Steve Abel, Rodrigo Alonso, Yunji Ha, Joerg Jaeckel, Nick Mavromatos, Arttu Rajantie and Michael Spannowsky for useful discussions.
This work is supported by the STFC under consolidated grant ST/P001246/1.

%\bigskip
%\newpage 
%\appendix
%\appendixpage
%\addappheadtotoc

\bigskip

%\newpage

\bibliographystyle{inspire}
\bibliography{main}

\clearpage
\appendix
\include{variables}

\end{document}